\documentclass[nohyper]{JHEP3}

\usepackage{url}
\usepackage{epsfig}
\usepackage{latexsym}
\usepackage{amsmath}
\usepackage{lscape}
\def\mathswitch#1{\relax\ifmmode#1\else$#1$\fi}

\newcommand{\si}{\sigma}

\newcommand{\NLO}{\mathrm{NLO}}

\newcommand{\TeV}{\unskip\,\mathrm{TeV}}

\def\nn{\nonumber}

\def\text{\textstyle}

\def\bc{\begin{center}}
\def\ec{\end{center}}
\def\bi{\begin{itemize}}
\def\ei{\end{itemize}}

\title{Supersymmetric top and bottom squark production at hadron
  colliders}

\author{Wim Beenakker\\
Theoretical High Energy Physics, IMAPP, Radboud University Nijmegen,\\
 P.O. Box 9010, NL-6500 GL Nijmegen, The Netherlands}
\author{Silja Brensing\\
  Institut f\"ur Theoretische Teilchenphysik und Kosmologie, RWTH Aachen University,\\
  D-52056 Aachen, Germany\\  
Nikhef Theory Group, Science Park 105, 1098 XG Amsterdam, The
 Netherlands}
\author{Michael Kr\"amer\\
  CERN, Physics Department, Theory Unit, CH-1211 Geneva 23, Switzerland\\
  Institut f\"ur Theoretische Teilchenphysik und Kosmologie, RWTH Aachen University,\\
  D-52056 Aachen, Germany}
  \author{Anna Kulesza\\
  Institut f\"ur Theoretische Teilchenphysik und Kosmologie, RWTH Aachen University,\\
  D-52056 Aachen, Germany}
\author{Eric Laenen\\
  ITFA, University of Amsterdam, Science Park 904, 1018 XE Amsterdam, \\
  ITF, Utrecht University, Leuvenlaan 4, 3584 CE Utrecht,\\
  Nikhef Theory Group, Science Park 105, 1098 XG Amsterdam, The
  Netherlands}
\author{ Irene Niessen\\
Theoretical High Energy Physics, IMAPP, Radboud University Nijmegen,\\
 P.O. Box 9010, NL-6500 GL Nijmegen, The Netherlands}

\abstract{The scalar partners of top and bottom quarks are expected to
  be the lightest squarks in supersymmetric theories, with 
  potentially large cross sections at hadron colli\-ders. We present
  predictions for the production of top and bottom squarks at the
  Tevatron and the LHC, including next-to-leading order corrections in
  supersymmetric QCD and the resummation of soft gluon emission at
  next-to-leading-logarithmic accuracy.  We discuss the impact of the
  higher-order corrections on total cross sections and transverse-momentum 
  distributions, and provide an estimate of the 
  theoretical uncertainty due to scale variation and the parton
  distribution functions.}


\preprint{\tiny CERN-PH-TH/2010-142\\[-2mm]
\tiny  ITFA-2010-015\\[-2mm]
\tiny  ITP-UU-10/17\\[-2mm]
\tiny  NIKHEF/2010-016\\[-2mm]
\tiny  TTK-10-33}

\begin{document}

\section{Introduction}
\label{sec:intro}
The search for supersymmetry~\cite{Golfand:1971iw, Wess:1974tw} is
a central part of the physics program at the proton--antiproton
collider Tevatron with a centre-of-mass energy of $\sqrt{S}=1.96$~TeV
and at the proton--proton collider LHC, which started operation in 2010 at
$\sqrt{S}=7$~TeV.  In particular squarks and gluinos, the
coloured supersymmetric particles, may be produced copiously in
hadronic collisions. The hadroproduction of top squarks
(stops)~\cite{Beenakker:1997ut} is an important special case, since
the strong Yukawa coupling between top quarks, stops and Higgs fields
gives rise to potentially large mixing effects and mass
splitting~\cite{Ellis:1983ed}. The same holds, albeit to a lesser
extent, for bottom squarks (sbottoms). Moreover, if the scalar masses
in unified supersymmetric theories are evolved from universal values
at high scales down to low scales, the lighter of the stop mass
eigenstates is generally driven to the lowest value in the entire squark mass
spectrum. The search for the lightest stop therefore plays a special role in
the quest to find signals of supersymmetry at hadron colliders.

Searches at LEP~\cite{LEP-SUSY, Heister:2002hp} and the
Tevatron~\cite{Abazov:2006fe}\,--\,\cite{Aaltonen:2010dy}
have placed lower limits on the lighter stop and sbottom mass eigenstates
in the range between about 70\,--\,200~GeV, depending on the
choice of supersymmetric parameters. The LHC will extend the range of
sensitivity into the TeV-region~\cite{Aad:2009wy,Bayatian:2006zz}.

In the minimal supersymmetric extension of the Standard Model (MSSM)
\cite{Nilles:1983ge, Haber:1984rc} with R-parity conservation, stops
are pair-produced at hadron colliders:
\begin{equation}
  pp/p\bar{p} \;\to\; \tilde{t}_1\bar{\tilde{t}}_1\, + X\quad \mbox{and}\quad
\tilde{t}_2\bar{\tilde{t}}_2\, + X \label{eq:processes}~,
\end{equation}
where $\tilde{t}_1$ and $\tilde{t}_2$ denote the lighter and heavier mass 
eigenstate, respectively. The hadroproduction of mixed 
$\tilde{t}_1\bar{\tilde{t}}_2$ or $\tilde{t}_2\bar{\tilde{t}}_1$ final states
is strongly suppressed since it can only proceed through electroweak channels 
or QCD-induced loop diagrams~\cite{Beenakker:1997ut, Berdine:2005tz, 
  Bozzi:2005sy}. Sbottom hadroproduction is described in a completely analogous
manner, so we will focus our discussion on stops.  We will, however,
comment on potential differences between stop and sbottom
hadroproduction, and provide benchmark cross sections for sbottom
production at the Tevatron and the LHC.

Accurate theoretical predictions for the stop-pair cross sections are crucial 
to derive exclusion limits~\cite{Abazov:2006fe}\,--\,\cite{Aaltonen:2010dy} 
and, in the case of discovery, can be used to
determine the stop masses and properties (see e.g.~Refs.~\cite{Kane:2008kw,
  Hubisz:2008gg, Dreiner:2010gv}). The cross sections for the stop-pair 
production processes (\ref{eq:processes}) have been calculated at 
next-to-leading order (NLO) in supersymmetric QCD 
(SUSY-QCD)~\cite{Beenakker:1997ut}. The SUSY-QCD corrections significantly 
reduce the renormalization- and factorization-scale dependence and increase 
the cross section with respect to the leading-order (LO) predictions if the 
renormalization and factorization scales are chosen
close to the stop mass. Electroweak corrections have been 
calculated as well~\cite{Hollik:2007wf, 
  Beccaria:2008mi}. Although they can be sizeable at large invariant masses and large transverse momenta,
they are moderate for the inclusive stop cross section. The SUSY-QCD calculation of 
Ref.~\cite{Beenakker:1997ut} has been implemented in the public computer code 
{\tt Prospino}~\cite{prospino} and presently forms the theoretical basis for
the stop mass limits obtained at the Tevatron. 

A significant part of the large NLO SUSY-QCD corrections to squark
hadroproduction can in general be attributed to the threshold 
region~\cite{Beenakker:1996ch, Beenakker:1997ut} where the partonic 
centre-of-mass energy is close to the kinematic threshold for producing massive
particles. In this region the NLO corrections are dominated by 
contributions from soft gluon emission off the coloured particles in
the initial and final state and by Coulomb corrections due to the
exchange of gluons between the massive particles in the final state.
The soft-gluon corrections can be taken into account to all orders in
perturbation theory by means of threshold resummation. A considerable
amount of work has recently been devoted to the calculation of threshold 
logarithms for total gluino and squark cross 
sections~\cite{Kulesza:2008jb}\,--\,\cite{Beenakker:2009ha}.
Final-state stops are excluded in these calculations and all other squark
flavours, the so-called light-flavour squarks, are treated as being mass 
degenerate, neglecting possible mixing effects. 
  
In this work, we extend the previous analyses of threshold resummation for the
hadro\-production of gluinos and mass-degenerate light-flavour squarks at 
next-to-leading logarithmic (NLL)
accuracy~\cite{Kulesza:2008jb,Kulesza:2009kq, Beenakker:2009ha}. Firstly, 
we consider the hadroproduction of stops and non-mass-degenerate 
sbottoms. Secondly, we study the impact of NLO and NLL corrections on the 
transverse-momentum distributions. Since theoretical predictions for 
differential distributions are input to the experimental analyses, it is 
important to assess how the shape of the distributions is affected by 
higher-order corrections. The threshold resummation for transverse-momentum 
distributions has been studied extensively for Standard-Model processes, see
e.g.~Refs.~\cite{Catani:1998tm}\,--\,\cite{Banfi:2004xa},
but not yet for SUSY processes.

The paper is structured as follows. In section~\ref{sec:stop_prod} we review 
the basic features of stop and sbottom hadroproduction, and we briefly discuss
the application of threshold resummation to the transverse-momentum 
distribution. Section~\ref{sec:stop_prod} also contains the specific stop-pair 
formulae that enter the calculation of the resummed cross sections. 
State-of-the-art SUSY-QCD predictions for stop hadroproduction at the
Tevatron and the LHC, including NLO corrections and NLL threshold
resummation, are presented in section~\ref{se:numres}. We discuss
the impact of the NLO+NLL corrections on total cross sections and
transverse-momentum distributions and provide an estimate of the
theoretical uncertainty due to scale variation and the
parton distribution functions. We conclude in section~\ref{se:conclusion}. 
The dependence of the stop and sbottom cross sections on the choice of 
supersymmetric parameters can be found in the appendix, where also some
predictions for specific benchmark scenarios are given.

\section{Stop and sbottom pair production}
\label{sec:stop_prod}

Let us first review some basic features of the stop and sbottom 
pair-production cross sections. At LO the hadroproduction of 
stop pairs proceeds through quark-antiquark annihilation and gluon-gluon 
fusion:
\begin{eqnarray}\label{eq:LO}
q\bar{q} & \to & \tilde{t}_1\bar{\tilde{t}}_1 \quad {\rm and} \quad \tilde{t}_2\bar{\tilde{t}}_2~,\nonumber\\
gg & \to & \tilde{t}_1\bar{\tilde{t}}_1 \quad {\rm and} \quad \tilde{t}_2\bar{\tilde{t}}_2~.
\end{eqnarray}
The corresponding Feynman diagrams are shown in Fig.~\ref{fig:feyn}. 
\vspace*{8mm}
\FIGURE{
\epsfig{file=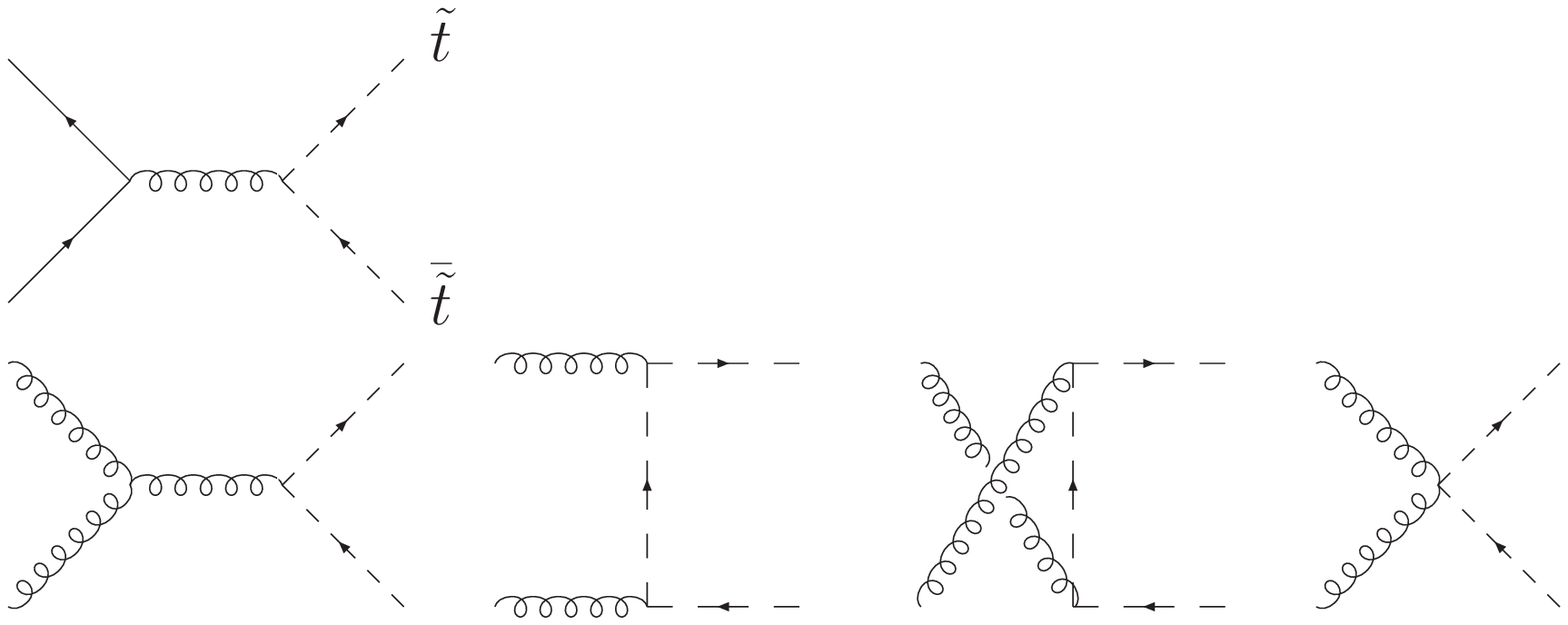, width=0.85\columnwidth}
\vspace*{1cm}
\caption{Leading-order Feynman diagrams for the production of stop pairs 
         through quark-antiquark annihilation (first line) and gluon-gluon 
         fusion (second line).}
\label{fig:feyn}}
In contrast to the hadroproduction of light-flavour squarks, no $t$-channel 
gluino-exchange graph occurs in the quark-antiquark channel. In such a 
$t$-channel graph the initial-state quarks should have the same flavour as the 
final-state squarks, but since top quarks are excluded as initial-state partons, 
the gluino-exchange graph is absent. 

The mass eigenstates $\tilde{t}_1$ and $\tilde{t}_2$ are
related to the weak interaction eigenstates $\tilde{t}_L$ and $\tilde{t}_R$
through mixing: $\tilde{t}_1 = \tilde{t}_L\cos\theta_{\tilde{t}} +
\tilde{t}_R\sin\theta_{\tilde{t}}\;$ and $\;\tilde{t}_2 =
-\,\tilde{t}_L\sin\theta_{\tilde{t}} + \tilde{t}_R\cos\theta_{\tilde{t}}$.  
The masses $m_{\tilde{t}_1}$, $m_{\tilde{t}_2}$ and the mixing angle 
$\theta_{\tilde{t}}$ are obtained from diagonalizing the stop mass matrix and are 
determined by Standard-Model and soft-supersymmetry-breaking 
parameters~\cite{Ellis:1983ed}. As mentioned in the introduction, 
mixed pairs $\tilde{t}_1\bar{\tilde{t}}_2$ or $\tilde{t}_2\bar{\tilde{t}}_1$ cannot 
be produced in lowest order QCD since 
the $g\tilde{t}\tilde{t}$ and $gg \tilde{t}\tilde{t}$ vertices are
diagonal in the chiral and in the mass basis.

The description of sbottom hadroproduction 
$pp/p\bar{p} \to \tilde{b}_1\bar{\tilde{b}}_1\,$ and
$\,\tilde{b}_2\bar{\tilde{b}}_2\,$ is completely analogous to that of stop 
production. The only differences occur in the 
$\,b\bar{b}\to \tilde{b}\bar{\tilde{b}}\,$ channel, where the initial-state 
bottom quarks do allow a $t$-channel gluino-exchange graph that gives rise to 
extra contributions. As will be discussed below, these contributions
lead to a less-suppressed LO threshold behaviour than the $s$-channel 
gluon-exchange contributions. However, we shall demonstrate in the appendix 
that their numerical impact on the hadronic cross sections is negligible. 
Thus, for all practical purposes, the LO and higher-order
cross-section predictions obtained for stop-pair production apply 
also to sbottom-pair production if the input parameters, i.e.~masses
and mixing angles, are modified accordingly.

When decomposed into $s$-channel colour states, the LO partonic cross sections
for the subprocesses (\ref{eq:LO}) read in generic notation:
\begin{align}
  \sigma^{(0)}_{q\bar q\to\tilde t\bar{\tilde t},{\bf 1}}
  &=\ 0~,\label{eq:LOqq1}\\[2mm]
  \sigma^{(0)}_{q\bar q\to\tilde t\bar{\tilde t},{\bf8}}
  &=\ \frac{\alpha_{\rm s}^2\pi(N_c^2-1)}{12N_c^2s}\,\beta^3~,\label{eq:LOqq8}
  \\[2mm]
  \sigma^{(0)}_{gg\to\tilde t\bar{\tilde t},{\bf1}}
  &=\ \frac{\alpha_{\rm s}^2\pi}{N_c(N_c^2-1)s}\,\biggl[ 
      \Bigl(\, \frac{1}{2} + \frac{2m^2}{s} \,\Bigr)\beta
      \,+ \Bigl(\, \frac{2m^2}{s} - \frac{4m^4}{s^2} \,\Bigr)
      \log\Bigl(\frac{1-\beta}{1+\beta}\Bigr) \biggr]~,\label{eq:LOgg1}\\[2mm]
  \sigma^{(0)}_{gg\to\tilde t\bar{\tilde t},{\bf8}}
  &=\ \frac{N_c^2-4}{2}\,\sigma^{(0)}_{gg\to\tilde t\bar{\tilde t},{\bf 1}}\nonumber\\[2mm]
     &\ \quad + \frac{\alpha_{\rm s}^2\pi N_c}{(N_c^2-1)s}\,\biggl[ 
      \Bigl(\, \frac{1}{12} + \frac{8m^2}{3s} \,\Bigr)\beta
      \, + \Bigl(\, \frac{m^2}{s} + \frac{2m^4}{s^2} \,\Bigr)
      \log\Bigl(\frac{1-\beta}{1+\beta}\Bigr) \biggr]~,\label{eq:LOgg8}
\end{align}
where $\alpha_{\rm s}$ is the strong coupling, $s$ the invariant partonic energy and
$N_c$ the number of colours. The colour labels $\mathbf{1}$ and $\mathbf{8}$ 
refer to the familiar singlet and octet colour states in SU(3), but all 
analytic results in this paper are derived for a general SU($N_c$)-theory in 
order to make the colour structure more transparent. Also note 
that we use a generic notation for the final-state particles and the associated
kinematics: $\tilde t$ can be either of the two stop mass eigenstates, with 
$m$ being the corresponding mass and $\beta = \sqrt{1-4m^2/s}\,$ the 
corresponding velocity. 

The expressions (\ref{eq:LOgg1}) and (\ref{eq:LOgg8}) for the gluon-gluon 
fusion cross section agree with the corresponding expression for 
light-flavour squark production, cf.\ Ref.~\cite{Beenakker:1996ch}. However,
the quark-antiquark annihilation contribution (\ref{eq:LOqq8}) is 
different due to the absence of the $t$-channel gluino-exchange graph. 
As a consequence, the LO cross section for 
$q\bar{q}\to\tilde{t}\bar{\tilde{t}}$ proceeds through $s$-channel gluon 
exchange only and is proportional to $\beta^3$, as opposed to $\beta$ for 
other squark flavours. This $\beta^3$ behaviour is the combined effect of the 
standard phase-space suppression factor $\beta$ and an additional $P$-wave 
suppression $\propto \beta^2$ near threshold: the pair of scalar particles 
needs to be produced in a $P$-wave state to balance the spin of the 
intermediate gluon.

Note that the LO cross sections (\ref{eq:LOqq1})--(\ref{eq:LOgg8}) only depend 
on the mass $m$ of the produced stops and not on any other supersymmetric 
parameters. At NLO, however, the stop mixing angle $\theta_{\tilde{t}}$ enters
through corrections involving the $t\tilde{t}\tilde{g}$ vertex and the 
four-squark couplings. As a result, already the analytical expressions for the 
$\tilde{t}_1\bar{\tilde{t}}_1$ and $\tilde{t}_2\bar{\tilde{t}}_2\,$ NLO cross sections 
are different.  Furthermore, virtual corrections involving squark and gluino loops introduce a
dependence of the stop-pair cross section on the masses of the squarks and 
the gluino. The dependence on all these other supersymmetric 
parameters turns out to be mild, as illustrated in the appendix.

In the threshold region the NLO stop-pair cross section is dominated by 
soft gluon emission, which leads to corrections $\propto \log^{i}\beta\,$ 
($i=1,2$), and Coulomb corrections $\propto 1/\beta\,$. In the notation of 
Ref.~\cite{Beenakker:1997ut} the threshold behaviour of the total NLO cross 
sections reads\footnote{Ref.~\cite{Beenakker:1997ut} contains a misprint 
   in the $\log(8{\beta^2})$-coefficient for the
   $q\bar{q}$-channel. In Eq.~(9) of~\cite{Beenakker:1997ut} the
   coefficient $-107/(36\pi^2)$ should be replaced by  $-155/(36\pi^2)$.}:
\begin{eqnarray}
  \sigma^{(1,\rm thr)}_{q\bar{q}\to \tilde{t}\bar{\tilde{t}}} 
  & = &
  \frac{\pi\alpha_{\rm s}^2(\mu^2)}{48m^2}\,\frac{N_c^2-1}{N_c^2}\,\beta^3
  \Biggl( 1 + 4\pi\alpha_{\rm s}(\mu^2)\,\biggl\{\,
          \frac{2C_F-C_A}{16\beta} \,-\, \frac{C_A}{4\pi^2}\log(8{\beta^2}) 
  \nonumber\\[1mm]
  && {}   +\, \frac{2C_F}{4\pi^2}\biggl[\,\log^2(8{\beta^2})
              - \frac{16}{3}\log(8{\beta^2})
              - \log(8{\beta^2})\log\Bigl(\frac{\mu^2}{m^2}\Bigr)\,\biggr]
                                         \biggr\}\Biggr)\label{eq:NLOqq}
\end{eqnarray}
and
\begin{eqnarray}
  \sigma^{(1,\rm thr)}_{gg\to \tilde{t}\bar{\tilde{t}}}
  & = & 
  \frac{\pi\alpha_{\rm s}^2(\mu^2)}{16m^2}\,\frac{N_c^2-2}{N_c(N_c^2\!-\!1)}\,\beta\,
  \Biggl( 1 + 4\pi\alpha_{\rm s}(\mu^2)\,\biggl\{\,
          \frac{2C_F-\frac{N_c^2-4}{N_c^2-2}\,C_A}{16\beta} 
          \,-\, \frac{N_c^2-4}{N_c^2-2}\,\frac{C_A}{4\pi^2}\log(8{\beta^2}) 
  \nonumber\\[1mm]
  && {}   +\, \frac{2C_A}{4\pi^2}\biggl[\,\log^2(8{\beta^2}) -4\log(8{\beta^2})
              - \log(8{\beta^2})\log\Bigl(\frac{\mu^2}{m^2}\Bigr)\,\biggr]
                                         \biggr\}\Biggr)\,.\label{eq:NLOgg}
\end{eqnarray}
Here $\mu$ denotes the factorization and renormalization scales, which we keep 
equal in this analysis. 
The first line in both expressions contains the gluon corrections that
can be attributed to the final-state particles, while the terms in the second line, 
proportional to either $ C_F \ =\ (N_c^2-1)/(2N_c)$ or $C_A \ =\ N_c$ correspond 
to initial-state radiation from a quark or a gluon line, respectively. 
The final-state radiation consists of two parts. The logarithmic soft-emission
terms are proportional to the colour charge of the 
final state~\cite{Bonciani:1998vc, Kulesza:2008jb, Beneke:2009rj, Beenakker:2009ha} and 
are therefore absent for the singlet colour state and proportional to
$C_A$ for the octet colour state. As a result, in the expression
for the gluon-gluon fusion channel these terms are multiplied by
$(N_c^2-4)/(N_c^2-2)$, which is the ratio of the colour-octet and the
total cross section at threshold (cf.~\cite{Bonciani:1998vc}). The Coulomb 
corrections receive contributions with colour factors $C_F$ and the 
total colour charge of the final state. 
Note, finally, that the threshold behaviour of the cross section is 
determined by QCD dynamics and does not involve any supersymmetric parameters
other than the mass of the produced stops.
       
\subsection{Threshold resummation for the total inclusive cross section}

In this paper we shall improve the NLO prediction for the hadroproduction of 
stop and sbottom pairs~\cite{Beenakker:1997ut} by resumming the leading and 
next-to-leading threshold logarithms of the form 
\begin{equation}
\alpha_{\rm s}^n \log^m\!\beta^2\ \ , \ \ m\leq 2n 
\label{eq:logbeta:structure}
\end{equation}
to all orders, and to next-to-leading logarithmic (NLL) accuracy.
The resummation for\-ma\-lism developed for heavy-quark 
production~\cite{Kidonakis:1997gm, Bonciani:1998vc, Kidonakis:2001nj} can be 
applied directly to squark-antisquark production as the colour structure is 
identical for both cases. The resummation is performed in the space of Mellin 
moments ($N$-space) following the procedure outlined in 
Refs.~\cite{Kulesza:2008jb,Kulesza:2009kq, Beenakker:2009ha}, where the 
hadroproduction of light-flavour squarks and gluinos at NLL accuracy has been 
considered. 

From the viewpoint of threshold resummation there is a notable
difference between the case of stops and light-flavour squarks, arising from 
the extra $\beta^2$ suppression of the quark-antiquark annihilation cross 
section $\sigma_{q\bar{q}\to \tilde{t}\bar{\tilde{t}}}\,$ near threshold. 
In $N$-space this effectively produces an extra factor $1/N$ compared to 
the light-flavour squark case, resulting in leading contributions 
$\propto\log^i(N)/N$ instead of $\log^i(N)$. For the seemingly 
analogous case of threshold resummation for the deep-inelastic structure 
function $F_L$, which has a leading behaviour of the type $\log^i(N)/N$ as 
well, differences from the expected NLL resummation structure were revealed in
Refs.~\cite{Moch:2009mu} and \cite{Grunberg:2009am}. However, the 
NLL resummation procedure developed for light-flavour squarks applies to
stop production as well. In the case of $F_L$ the 
extra factor $1/N$ with respect to the $N$ dependence of the structure
function $F_2$ arises due to a special initial-state jet function 
\cite{Akhoury:1998gs,Akhoury:2003fw} associated with the longitudinal 
projection. In contrast, in our case the $\log^i(N)/N$ dependence arises from 
projecting onto the $P$-wave final state, which does not depend on the
initial state jet function. Moreover, at 
$\mathcal{O}(\alpha_{\rm s})$ leading and subleading $\log(N)/N$ corrections 
can be computed from the one-loop calculations, and they do exhibit the pattern 
expected from NLL threshold resummation.
Finally, in view of the different threshold behaviour of the $q\bar{q}$ and 
$gg$ channels one might worry about the possibility that these channels mix in 
the NLL threshold resummation as a result of soft-quark emissions, which is not
the case for top-quark and light-flavour squark production. However, we have 
checked that to NLL accuracy  no such mixing occurs at $\mathcal{O}(\beta^3)$.
Based on these observations we are confident that the expressions of 
Refs.~\cite{Kulesza:2008jb,Kulesza:2009kq,Beenakker:2009ha} can be applied to
inclusive stop-pair production as well.

The new elements that enter the NLL resummed cross section for 
stop pair production are the LO partonic cross sections in $N$-space
(indicated by a tilde), decomposed into $s$-channel singlet and octet colour 
states. They are given by:
\begin{eqnarray}
\label{eq:LO-Nspace}
  \hspace*{-6.5ex}
  \tilde\sigma_{q\bar{q}\to \tilde{t}\bar{\tilde{t}},\mathbf{1}}^{(0)}(N)
  &=& 0~, \\[1mm]
  \hspace*{-6.5ex}
  \tilde\sigma_{q\bar{q}\to \tilde{t}\bar{\tilde{t}},\mathbf{8}}^{(0)}(N)
  &=& \frac{\alpha_{\rm s}^2\pi^{3/2}}{{64m^2}}\,
      \frac{N_c^2-1}{N_c^2}\,\frac{\Gamma(N+1)}{\Gamma(N+7/2)}~,\\[1mm]
  \hspace*{-6.5ex}
  \tilde\sigma_{gg\to \tilde{t}\bar{\tilde{t}},\mathbf{1}}^{(0)}(N)
  &=& \frac{\alpha_{\rm s}^2\pi^{3/2}}{{16m^2}}\,
      \frac{1}{N_c(N_c^2-1)}\,\frac{N(N+3)+4}{(N+2)(N+3)}\,
      \frac{\Gamma(N+1)}{\Gamma(N+5/2)}~,\label{eq:ggN1} \\[1mm]
  \hspace*{-6.5ex}
  \tilde\sigma_{gg\to \tilde{t}\bar{\tilde{t}},\mathbf{8}}^{(0)}(N)
  &=& \frac{N_c^2-4}{2}\,
      \sigma_{gg\to \tilde{t}\bar{\tilde{t}},\mathbf{1}}^{(0)}(N)
      \,+\, \frac{\alpha_{\rm s}^2\pi^{3/2}}{{64m^2}}\,
            \frac{N_c}{N_c^2-1}\,\frac{N(N+3)+4}{(N+2)(N+3)}\,
            \frac{\Gamma(N+1)}{\Gamma(N+7/2)}\,. \label{eq:ggN8}
\end{eqnarray}
The results for the gluon-gluon channels (\ref{eq:ggN1}) and (\ref{eq:ggN8}) 
agree with those presented in \cite{Kulesza:2009kq} for light-flavour squarks 
and $N_c=3$, and are reproduced here for completeness for general SU$(N_c)$. 
Note that the results in \cite{Kulesza:2009kq} include an additional factor 
$2n_f$ from the summation over the $L$ and $R$~squarks of $n_f$ light 
flavours.

\subsection{Threshold resummation for the transverse-momentum distribution}

Similar to the inclusive cross section, soft-gluon corrections can also 
dominate the transverse-momentum distribution of the stops if the production 
takes place sufficiently close to the edge of phase space. We now briefly
review the general construction of threshold-resummed transverse-momentum 
distributions (cf.\ \cite{Catani:1998tm}-\cite{Banfi:2004xa}) and its application to 
stop-pair production.

We start with the general framework applicable to the hadroproduction of a 
pair of massive coloured particles. Assuming that one of the final-state 
particles carries a transverse momentum $p_T$, the minimal energy necessary to
produce the system is $2m_T$, where the transverse mass $m_T$ is defined by
\begin{equation}
  m_T \ =\ \sqrt{m^2+p_T^2}~.
\end{equation}
For the $p_T$-distribution the dominant contributions originating from soft 
gluon emission have again the structure (\ref{eq:logbeta:structure}), with the
variable $\beta$ replaced by (cf.~\cite{Laenen:1991af})
\begin{equation}
  \beta_T \ =\ \sqrt{1-4 m_T^2/s}~.
\end{equation}
Like in the case of the total inclusive cross section, the resummation of the 
logarithmic threshold corrections to $p_T$-distributions takes place in the 
space of Mellin moments. However, in this case the Mellin transform is taken 
with respect to the scaling variable $\hat{x}_T^2=4m_T^2/s$, i.e. 
\begin{equation}
  \frac{d {\tilde{\sigma}_{ij\rightarrow kl}}}{d p_T}
  \bigl(N,p_T,\mu^2\bigr) \equiv \int_0^1 d \hat {x}_T^2\,(\hat {x}_T^2)^{N-1}
  \,\frac{d \sigma_{ij\rightarrow kl}}{d p_T}\bigl(\hat x_T^2,p_T,\mu^2\bigr)
  \label{eq:3}
\end{equation}
for a generic partonic subprocess $\,ij\to kl$.

In $N$-moment space the resummed partonic $p_T$-distribution for the 
hadroproduction of two massive coloured particles is given by
\begin{eqnarray}
  \label{eq:12}
  \frac{d {\tilde{\sigma}_{ij\rightarrow kl}}^{\rm (res)}}{d p_T} 
  \bigl(N,p_T,\mu^2\bigr) 
  &=& \sum_{I}\,\frac{d {\tilde{\sigma}_{ij\rightarrow kl,I}}^{(0)}}{d p_T} 
      \bigl(N,p_T,\mu^2\bigr)\,C_{ij \rightarrow kl, I}\bigl(N,p_T,\mu^2\bigr) 
      \\[1mm]
  && \times\,\Delta_i (N\!+\!1,Q^2,\mu^2)\,\Delta_j (N\!+\!1,Q^2,\mu^2)\,
     \Delta^{\rm (s)}_{ij\rightarrow kl,I}\bigl(N+1,Q^2,\mu^2\bigr)~, \nn
\end{eqnarray}
with $Q=2 m_T$ representing the resummation scale and the index $I$ indicating
all possible colour states of the hard scattering. To NLL accuracy, the 
coefficients $C_{ij \rightarrow kl, I}\bigl(N,p_T,\mu^2\bigr)$ are equal to $1$
for all channels and colour structures. The functions $\Delta_i$ and $\Delta_j$
sum the effects of the (soft-)collinear radiation from the incoming partons. 
They are process- and colour-independent, and are therefore a universal 
ingredient in all threshold-resummed cross sections. Explicit expressions for
the $\Delta_i$ can for instance be found in Ref.~\cite{Kulesza:2009kq}. 
The function $\Delta^{\rm(s)}_{ij\rightarrow kl,I}$
accounts for soft wide-angle emission and depends on the production process 
and the colour channel. At NLL accuracy it is given by
\begin{equation}
  \Delta_{ij\rightarrow kl,I}^{\rm (s)}\bigl(N,Q^2,\mu^2 \bigr) 
  \;=\; \exp\Big[\int_{\mu}^{Q/N}\frac{dq}{q}\,\frac{\alpha_{\rm s}(q)}{\pi}
                 \,D_{ij\rightarrow kl,I} \,\Big]\,.
\label{eq:1}
\end{equation}

The values of the coefficients $D_{ij\rightarrow kl,I}$ follow from the
threshold limit of the one-loop soft anomalous-dimension matrix in
the way described e.g.~in Ref.~\cite{Beenakker:2009ha}. If an $s$-channel 
colour basis is used, the soft anomalous-dimension matrix for the  
$p_T$-distribution becomes diagonal in colour space in the threshold limit
$\beta_T \to 0$, leading to the colour-diagonal form of the resummation 
formula given in Eq.~(\ref{eq:12}). This is similar to threshold resummation 
for the total cross section, where the soft anomalous-dimension matrix becomes
diagonal in colour space in the corresponding threshold limit $\beta \to 0$.
Since the threshold limit is defined differently for the total inclusive cross
section and the $p_T$-distribution, the values of the 
$D$-coefficients are different as well; in particular, the 
$D$-coefficients now carry $p_T$-dependence.

For the stop-pair production processes we have
\begin{eqnarray}
  D_{q\bar{q}/gg\to \tilde{t}\bar{\tilde{t}},\mathbf{1}} 
  &=& -\,2C_F\,\bigl( \mathrm{Re}\,L_{\bar\beta_T} + 1 \bigr)~,\\[3mm]
  D_{q\bar{q}/gg\to \tilde{t}\bar{\tilde{t}},\mathbf{8}} 
  &=& -\,2C_F\,\bigl( \mathrm{Re}\,L_{\bar\beta_T} + 1 \bigr)
      \,+\, C_A\biggl[ \log\Bigl(\frac{m_T^2}{m^2}\Bigr) 
                       + \mathrm{Re}\,L_{\bar\beta_T} \biggr]\,,
  \label{eq:D-coeffs-pt}
\end{eqnarray}
with 
\begin{equation}
  \label{eq:L-beta-bar}
  \mathrm{Re}\,L_{\bar\beta_T} 
  \ =\ \frac{1+{\bar\beta_T}^2}{2{\bar\beta_T}}
       \log\Bigl( \frac{1-{\bar\beta_T}}{1+{\bar\beta_T}} \Bigr)
\end{equation}
and 
\begin{equation}
  \label{eq:beta-bar}
  {\bar\beta_T} \;= \lim_{\hat{x}_T^2 \to 1}\beta_T 
  \;=\; \sqrt{1-{m^2}/{m_T^2}}~.
\end{equation}
These expressions agree with those obtained for heavy-quark production in
the context of joint resummation~\cite{Banfi:2004xa}.

In addition to the soft radiative factor, the other new elements which
have to be calculated in order to obtain resummed predictions from
Eq.~(\ref{eq:12}) are the colour-decomposed LO
\mbox{$p_{T}$-distributions} in $N$-moment space. They are obtained from
\begin{eqnarray}
\label{eq:LO-pt-xspace1}
  \frac{d\sigma_{q\bar{q}\to \tilde{t}\bar{\tilde{t}},\mathbf{1}}^{(0)}}
       {d p_{T}}
  &=& 0~, \\[1mm]
  \frac{d\sigma_{q\bar{q}\to \tilde{t}\bar{\tilde{t}},\mathbf{8}}^{(0)}}
       {d p_{T}}
  &=& \frac{\alpha_{\rm s}^2 \pi}{s^3}\,\frac{2(N_c^2-1)}{N_c^2}\,
      \frac{p_{T}^3}{\beta_{T}}~, \\[1mm]
  \frac{d \sigma_{gg\to \tilde{t}\bar{\tilde{t}},\mathbf{1}}^{(0)}}{d p_{T}}
  &=& \frac{\alpha_{\rm s}^2 \pi}{s^2}\,\frac{2}{N_c (N_c^2-1)}\,
      \frac{p_{T}}{\beta_{T}}\,\frac{m^4+p_{T}^4}{m_{T}^4}~, \\[1mm]
  \frac{d \sigma_{gg\to \tilde{t}\bar{\tilde{t}},\mathbf{8}}^{(0)}}{d p_{T}}
  &=& \frac{N_c^2-4}{2}\,
      \frac{d\sigma_{gg\to \tilde{t}\bar{\tilde{t}},\mathbf{1}}^{(0)}}{d p_{T}}
      \,+\, \frac{\alpha_{\rm s}^2\pi}{s^2}\,\frac{N_c}{N_c^2-1}\,p_{T}\beta_{T}\,
            \frac{{m^4}+p_{T}^4}{m_{T}^4}\,.\label{eq:LO-pt-xspace4}
\end{eqnarray}
Taking a Mellin transform of~(\ref{eq:LO-pt-xspace1})-(\ref{eq:LO-pt-xspace4})  with respect to the threshold variable $\hat{x}_T^2$
one finds
\begin{eqnarray}
\label{eq:LO-pt-Nspace}
  \hspace*{-6ex}
  \frac{d\tilde{\sigma}_{q\bar{q}\to\tilde{t}\bar{\tilde{t}},\mathbf{1}}^{(0)}}
       {d p_{T}}(N)
  &=& 0~,\\[1mm]
  \hspace*{-6ex}
  \frac{d\tilde{\sigma}_{q\bar{q}\to\tilde{t}\bar{\tilde{t}},\mathbf{8}}^{(0)}}
       {d p_{T}}(N)
  &=& \frac{\alpha_{\rm s}^2\pi^{3/2}}{32}\,\frac{N_c^2-1}{N_c^2}\,
      \frac{\Gamma(N+3)}{\Gamma(N+7/2)}\,\frac{p_{T}^3}{m_{T}^6}~,
      \\[1mm]
  \hspace*{-6ex}
  \frac{d\tilde{\sigma}_{gg\to\tilde{t}\bar{\tilde{t}},\mathbf{1}}^{(0)}}
       {d p_{T}}(N)
  &=& \frac{\alpha_{\rm s}^2\pi^{3/2}}{8}\,\frac{1}{N_c(N_c^2-1)}\,
      \frac{\Gamma(N+2)}{\Gamma(N+5/2)}\,\frac{p_{T}({m^4}+p_{T}^4)}{m_{T}^8}~,
     \\[1mm]
  \hspace*{-6ex}
  \frac{d\tilde{\sigma}_{gg\to\tilde{t}\bar{\tilde{t}},\mathbf{8}}^{(0)}}
       {d p_{T}}(N)
  &=& \frac{N_c^2-4}{2}\,
      \frac{d\tilde{\sigma}_{gg\to\tilde{t}\bar{\tilde{t}},\mathbf{1}}^{(0)}}
           {d p_{T}}(N) 
      \,+\, \frac{\alpha_{\rm s}^2\pi^{3/2}}{32}\,\frac{N_c}{N_c^2-1}\,
            \frac{\Gamma(N+2)}{\Gamma(N+7/2)}\,
            \frac{p_{T}({m^4}\!+p_{T}^4)}{m_{T}^8}\,.
\end{eqnarray}

Having calculated the resummed partonic expression in $N$-space,
Eq.~(\ref{eq:12}), the resummed hadronic $p_T$-distribution is obtained by
the inverse Mellin transform \begin{eqnarray}
  \label{eq:14}
  \nn
  \frac{d \si^{\rm (res)}_{h_1 h_2 \to kl}}{d p_T}\bigl(x_T^2, p_T,\mu^2\bigr)
  \ = \sum_{i,j=q,\bar{q},g}\,\int_\mathrm{CT}\,\frac{dN}{2\pi i}\,
(x_T^2)^{-N}
  && \hspace*{-1.8ex}\tilde f_{i/h_1}(N\!+\!1,\mu^2)\,
     \tilde f_{j/h_{2}}(N\!+\!1,\mu^2) \\
  && \hspace*{-1.8ex}\times\,
     \frac{d {\tilde{\sigma}_{ij\rightarrow kl}}^{\rm (res)}}
          {d p_T}\bigl(N, p_T,\mu^2\bigr)~, \end{eqnarray} where
$x_T^2=4m_T^2/S$ is the hadronic scaling variable. The functions $\tilde
f_{i/h_1}\,$ and  $\,\tilde f_{j/h_2}$ are the Mellin moments of the parton
distribution functions for the initial-state hadrons $h_1$ and $h_2$.
In order to retain the information contained in the NLO
$p_T$-distributions~\cite{Beenakker:1997ut}, the NLO and NLL results are
combined through a matching procedure that avoids double counting of the
logarithmic terms in the following way:
\begin{eqnarray}
  \label{eq:15}
  \frac{d \si^{\rm (NLO+NLL)}_{h_1 h_2 \to kl}}{d p_T}\,
  \bigl(x_T^2,p_T,\mu^2\bigr)
  \;&=&\; \frac{d {{\sigma}_{h_1 h_2\to kl}}^{\rm (NLO)}}{d p_T}\,
          \bigl(x_T^2, p_T,\mu^2\bigr) \nn\\[1mm]
    && \hspace*{-10ex} +\, \sum_{i,j=q,\bar{q},g}\,\int_\mathrm{CT}\,
       \frac{dN}{2\pi i}\,(x_T^2)^{-N}\,
       \tilde f_{i/h_1}(N+1,\mu^2)\,\tilde f_{j/h_{2}}(N+1,\mu^2) \nn\\[3mm]
    && \hspace*{-10ex}\times\,
       \left[\frac{d {\tilde{\sigma}_{ij\rightarrow kl}}^{\rm (res)}}{d p_T}\,
             \bigl(N, p_T,\mu^2\bigr)
       \,-\, \frac{d {\tilde{\sigma}_{ij\rightarrow kl}}^{\rm (res)}}{d p_T}\,
             \bigl(N, p_T,\mu^2\bigr)\biggr|_{\scriptscriptstyle({\NLO})}\, 
       \right] ,
\end{eqnarray} 
where $(d {\tilde{\sigma}}^{\rm (res)}/d p_T)
|_{\scriptscriptstyle({\NLO})}$ represents the perturbative expansion of
  the NLL $p_T$-distribution (\ref{eq:12}) truncated at the order of $\alpha_{\rm s}$ associated with NLO.
To evaluate the inverse Mellin transform in Eqs.~(\ref{eq:14}) and 
(\ref{eq:15}) we adopt the ``minimal prescription'' of
Ref.~\cite{Catani:1996yz} for the integration contour CT.

\section{Numerical results}
\label{se:numres}

We present numerical results for the NLL-resummed cross sections and
transverse-mo\-men\-tum distributions matched with the complete NLO results 
for stop-pair production at the Tevatron ($\sqrt{S}=1.96$~TeV) and the LHC
($\sqrt{S}=7$ and 14~TeV). The matching procedures for the total cross section 
and for the transverse-momentum distribution are described in 
Ref.~\cite{Beenakker:2009ha} and in Eq.~(\ref{eq:15}), respectively. We use the notation NLO+NLL 
for matched quantities in the following. The NLO corrections are calculated 
using {\tt Prospino}~\cite{prospino}, based on the calculation presented in 
Ref.~\cite{Beenakker:1997ut}. The QCD coupling $\alpha_{\rm s}$ and the parton 
distribution functions (pdfs) at NLO are defined in the $\overline{\rm MS}$ 
scheme with five active flavours. The mass of the stop is renormalized in the 
on-shell scheme and the SUSY particles are decoupled from the running of 
$\alpha_{\rm s}$ and the pdfs. Since mixing enters explicitly only through 
higher-order diagrams, the angle $\theta_{\tilde{t}}$ need not be renormalized and 
one can use the lowest-order expression derived from the stop mass matrix.

As our default, hadronic total cross sections and transverse-momentum 
distributions are obtained with the 2008 NLO MSTW pdfs~\cite{MSTW} and the 
corresponding $\alpha_{\rm s}(M_{Z}) = 0.120$. The NLL corrections are 
convoluted with pdfs in Mellin space, derived with the program 
{\tt PEGASUS}~\cite{Vogt:2004ns} based on the MSTW parametrization at the
initial factorization scale. For the total hadronic cross sections, we have 
also used the method of Ref.~\cite{Kulesza:2002rh} to evaluate the NLL cross 
section with standard parametrizations of pdfs in $x$-space.  We find however a 
much better numerical stability when the total NLL cross
sections are evaluated with Mellin-space pdfs. 

Beyond LO the cross section does not only depend on the stop mass, but also on 
the gluino mass $m_{\tilde{g}}$, the average mass of the first and second 
generation squarks $m_{\tilde{q}}$ and the mixing angle $\theta_{\tilde{t}}$. For this 
reason we have adopted the SPS1a' benchmark scenario~\cite{AguilarSaavedra:2005pw}
for our numerical analysis. Taking a top-quark mass of 
$m_t=172.5$ GeV~\cite{Amsler:2008zzb} and $\alpha_{\rm s}(M_{Z}) = 0.120$ the 
SPS1a' scenario 
corresponds to $m_{\tilde{g}} = 610$~GeV, $m_{\tilde{q}} = 560$~GeV,
$\sin(2\theta_{\tilde{t}}) = 0.932$ and stop masses of $m_{\tilde{t}_1} = 367$~GeV 
and $m_{\tilde{t}_2} = 590$~GeV~\cite{Porod:2003um}. However, in order
to focus on the mass dependence of the cross section and the NLO+NLL 
corrections, we vary the mass of the stop while keeping the other SUSY 
parameters fixed. As shown in Ref.~\cite{Beenakker:1997ut} and discussed in 
more detail in the appendix, the dependence of the cross section on the 
additional SUSY parameters is small, justifying this procedure. Note that the 
numerical results presented for stop production also apply to sbottom
production when the same input parameters are adopted. In the appendix we show 
that the impact of bottom-quark induced contributions to sbottom 
hadroproduction is negligible and present benchmark predictions for the sbottom
cross section.

\subsection{Results for the total cross section}

Let us first discuss the scale dependence of the SUSY-QCD total cross-section 
prediction. Fig.~\ref{fig:scale} shows the scale dependence in LO, NLO and 
NLO+NLL for $\tilde{t}_1\bar{\tilde{t}}_1$ production, using 
$m_{\tilde{t}_1} = 200$~GeV and $500$~GeV at the Tevatron and LHC, 
respectively. Here and in the following, we present results for the LHC 
operating at 7~TeV and at 14~TeV centre-of-mass energy. Note that the LO 
predictions are obtained with LO pdfs and the corresponding LO values for 
$\alpha_{\rm s}$~\cite{MSTW}. The renormalization and factorization scales are 
identified and varied around the central scale $\mu_0 = m_{\tilde{t}_1}$ from 
$\mu = \mu_0/10$ up to $\mu = 5\mu_0$. We observe the usual strong reduction 
of the scale dependence when going from LO to NLO. A further significant 
improvement is obtained when the resummation of threshold logarithms is 
included, in particular for stop production at the Tevatron and at the LHC 
running at 7~TeV. 

Near the central scale $\mu = m_{\tilde{t}_1}$ the cross section is enhanced 
by the SUSY-QCD corrections at NLO and NLO+NLL. The size of the 
$K$-factors $K_{\rm NLO} \equiv \sigma_{\rm NLO}/\sigma_{\rm LO}$ and 
$K_{\rm NLL} \equiv \sigma_{\rm NLO+NLL}/\sigma_{\rm NLO}$ strongly depends on
the stop mass and the collider, as is shown in Fig.~\ref{fig:K_NLO_K_NLL}. 
At the Tevatron, where the cross section is dominated by 
$q\bar{q}$-annihilation for large stop masses, the NLO $K$-factor is moderate 
and ranges from roughly~1.2~to~1.03 for stop masses in the range between
100~and~300~GeV. A further enhancement by 
up to 7\% is found for large stop masses when the NLL resummation is included. 
At the LHC, the $gg$ initial state is dominant and the QCD corrections are in 
general larger. For $\tilde{t}_1\bar{\tilde{t}}_1$ and 
$\tilde{t}_2\bar{\tilde{t}}_2$ production at the LHC we consider the mass 
ranges $100~{\rm GeV}\le m_{\tilde{t}_1} \le 550$~GeV (lower horizontal axis) 
and $550~{\rm GeV}\le m_{\tilde{t}_2} \le 1$~TeV (upper horizontal axis).
At 7~TeV we find NLO corrections ranging from about 40\% at the
lower end of the stop mass range to about 20\% for stop masses near 1~TeV.
The NLL resummation leads to a further increase of the cross-section 
prediction of approximately 10\% for stop masses in the TeV-range. At 14 TeV 
centre-of-mass energy the NLO corrections to stop production are significant 
and increase the LO cross section by around 35\% for moderate
$\tilde{t}_1$ masses and by up to 40\% for $\tilde{t}_2$ with $m_{\tilde{t}_2} \approx 600$~GeV, while 
the impact of the NLL resummation is modest with at most 5\% further 
increase for stop masses in the TeV-range. The singularities at the stop-decay 
threshold $m_{\tilde{t}} = m_t + m_{\tilde{g}} = 782.5$~GeV originate from the
stop wave-function renormalization. They are an unphysical artefact of the  
on-shell approach of Ref.~\cite{Beenakker:1997ut} and could be removed by taking 
into account the finite widths of the unstable stops. Note that the NLO cross sections for 
$\tilde{t}_1\bar{\tilde{t}}_1$ and $\tilde{t}_2\bar{\tilde{t}}_2$ production 
are not identical, even if the masses are taken equal. The reason for this is 
that the stop mixing angle contributes in different ways to both reactions at 
NLO, as discussed in section~\ref{sec:stop_prod}. Furthermore, while we 
vary the mass of the stop particle that appears in the final state, the mass 
of the other stop, which enters the loop corrections, is set to its SPS1a' value 
and thus differs for $\tilde{t}_1$ and $\tilde{t}_2$. However, numerically the difference 
between the two NLO total cross sections is moderate. The NLL resummation 
does not involve any SUSY parameters apart from the stop mass itself and
thus affects the $\tilde{t}_1\bar{\tilde{t}}_1$ and 
$\tilde{t}_2\bar{\tilde{t}}_2$ cross sections in the same way. 
The NLL $K$-factors have a tiny SUSY-parameter dependence, which enters through
the ratio $\sigma_{\rm NLO+NLL}/\sigma_{\rm NLO}$. 

Predictions for the LO, NLO, and NLO+NLL total cross sections are shown in 
Fig.~\ref{fig:total} and Tables~\ref{tab:tevatron}--\ref{tab:lhc14} for 
$\tilde{t}_1\bar{\tilde{t}}_1$ production at the Tevatron and 
$\tilde{t}_1\bar{\tilde{t}}_1$/$\tilde{t}_2\bar{\tilde{t}}_2$ production at 
the LHC with 7~TeV and 14~TeV centre-of-mass energy. In fact, the cross 
sections for $\tilde{t}_1\bar{\tilde{t}}_1$ and $\tilde{t}_2\bar{\tilde{t}}_2$
production at equal masses are indistinguishable on the scale of 
Fig.~\ref{fig:total}. We thus refrain from showing additional plots for 
$\tilde{t}_2\bar{\tilde{t}}_2$ production. The results shown in 
Fig.~\ref{fig:total} represent the state-of-the-art SUSY-QCD predictions at 
NLO+NLL accuracy. The error bands include the NLO+NLL scale variation in the 
range $m_{\tilde{t}_1}/2 \le \mu \le 2m_{\tilde{t}_1}$ as well as the NLO 
pdf uncertainty, added in quadrature. The pdf
uncertainty is obtained with the help of the 90\% C.L. MSTW error 
pdfs~\cite{Martin:2009iq}. More detailed information is available in
Tables~\ref{tab:tevatron}--\ref{tab:lhc14}. 

In Table~\ref{tab:tevatron} we present results for 
$\tilde{t}_1\bar{\tilde{t}}_1$ production at the Tevatron. As discussed before,
we observe an increase of the cross-section prediction near the central scale 
when going from LO to NLO and a further enhancement when NLL threshold 
resummation is included. The scale dependence in the range 
$m_{\tilde{t}_1}/2 \le \mu \le 2m_{\tilde{t}_1}$ is reduced from about 
$\pm 50\%$ at LO to about $\pm 10\%$ at NLO+NLL. The estimated pdf uncertainty
is approximately 5\%. We also present cross-section predictions obtained with 
the CTEQ6 pdf set~\cite{Nadolsky:2008zw} and an estimate of the corresponding 
pdf error. The difference between the MSTW and CTEQ results is particularly 
pronounced at large stop masses, $m_{\tilde{t}_1}\approx 300$ GeV, where the 
cross sections obtained with CTEQ pdfs are about 7\% larger than the ones 
obtained with MSTW pdfs. We observe that the CTEQ pdf error estimate of 
about 10\% is roughly twice as large as that of MSTW.

Results for $\tilde{t}_1\bar{\tilde{t}}_1$ and $\tilde{t}_2\bar{\tilde{t}}_2$ 
production at the LHC with 7~TeV are collected in Table~\ref{tab:lhc7}. 
Here the impact of the SUSY-QCD corrections at NLO is large, with the NLL 
resummation adding a further enhancement of up to 10\%. The scale uncertainty 
of the NLO+NLL prediction is reduced to a level of about 10\%. Unfortunately 
the pdf error is sizeable, in particular at stop masses in the 
TeV-region, where we find a pdf error of about 20\% from MSTW. Also the 
difference between MSTW and CTEQ is significant for large stop masses, with a 
25\% increase in the prediction for stop masses near 1~TeV when going from 
MSTW to CTEQ pdfs. As before, we find a pdf error from the CTEQ analysis that
is about twice as large as that of MSTW and reaches 45\% for 
$m_{\tilde{t}} \approx 1$~TeV. Of course, the large pdf uncertainty is not a 
specific feature of stop production. It rather generically affects predictions 
for TeV-scale particle production at the LHC with 7~TeV centre-of-mass energy, 
since these predictions are particularly sensitive to the gluon pdf at large 
$x$ (see e.g.~Ref.~\cite{Ball:2010de}). The conclusion therefore is that more 
accurate pdf determinations are needed in order to allow for a precise 
prediction of heavy-particle production during the initial phase of the LHC at 
7~TeV centre-of-mass energy. 

Going from 7~TeV to 14~TeV at the LHC, we observe in Table~\ref{tab:lhc14} a 
significant increase in the predicted cross section of about a factor of 4 for 
stop masses around 100~GeV and up to a factor of about 60 for masses in the 
TeV-region and MSTW pdfs. Just like at 7~TeV, the scale uncertainty of the 
NLO+NLL prediction is down to a level of about 10\%. The pdf uncertainty is 
more moderate than at 7~TeV, ranging from 3\% at small masses to about 10\% at 
large stop masses for MSTW, and correspondingly from 3\% to 20\% for CTEQ.

\subsection{Results for the transverse-momentum distribution}

Let us now turn to the discussion of the transverse-momentum distributions. 
Figure~\ref{fig:pt_m} shows a comparison between the LO, NLO and 
NLO+NLL distributions normalized to unity. We use normalized distributions in 
order to be able to directly read off the NLO- and NLL-induced changes in the 
shape of the distribution. As for the previous results, we have 
used the stop mass as the central scale, $\mu = m$, in Figure~\ref{fig:pt_m}. 
This is a possible choice as we do not consider regions where $p_T \gg m$ 
and where a $p_T$-dependent scale would have been mandatory.
As already observed in Ref.~\cite{Beenakker:1998wi}, the 
transverse momentum carried away by hard gluon radiation in higher orders 
softens the NLO transverse-momentum distribution with respect to the LO 
distribution. This effect is particularly visible at the Tevatron and at the 
LHC with 7~TeV centre-of-mass energy. The NLL soft-gluon resummation, on the 
other hand, does not affect the shape of the distribution significantly. 
To elucidate the impact of the higher-order corrections more clearly, we 
display the transverse-momentum dependence of the NLO and NLL $K$-factors in 
Figure~\ref{fig:pt_mt}, this time using the transverse mass 
$m_T = \sqrt{m^2 + p_T^2}$ as the scale. The significant softening of the 
transverse-momentum distribution at NLO at the Tevatron and the 7~TeV LHC is 
reflected in the variation of the $K$-factor, with $K_{\rm NLO}$ dropping from
roughly $1.8$ at small $p_T$ to a value near one at $p_T \approx 2m$. 
In comparison, the impact of the NLL resummation is small. A similar behaviour,
albeit less pronounced, is observed at the LHC with 14~TeV centre-of-mass
energy. It would be interesting to see if using NLO+NLL transverse-momentum 
distributions would affect the experimental analyses, which so far have been 
based on LO Monte Carlo predictions. In this context we recall that the shape 
of the stop rapidity distribution is not changed significantly by higher-order 
corrections, see Ref.~\cite{Beenakker:1998wi}. 

\section{Conclusions}
\label{se:conclusion}

In this paper we have performed the NLL threshold resummation for stop
and sbottom hadroproduction, considering both the inclusive cross sections and
the transverse-momentum distributions. As the lighter stop and sbottom mass
eigenstates are generally predicted to be the lightest strongly interacting 
SUSY particles, the search for these particles plays a special role in the 
quest to find signals of supersymmetry at hadron colliders.  

Results have been given for the Tevatron and for the LHC running at both
7 TeV and 14 TeV centre-of-mass energy. Compared to the NLO predictions for the total cross section, the 
NLL corrections lead to a significant reduction of the scale dependence
and increase the cross section by up to 10\% for masses in the TeV range if 
the renormalization and factorization scales are chosen close to the mass of
the final-state particles. We have also studied the SUSY parameter 
dependence of the stop and sbottom cross sections and find small variations 
of at most 2\%. The size of bottom-induced contributions to sbottom pair production 
is negligible numerically so that predictions obtained for stop-pair production 
also apply to sbottom-pair production when the same input parameters 
are adopted.

Since $p_T$ cuts are used extensively in experimental analyses, which at 
present are based on LO Monte Carlo simulations, it is important to investigate 
how the NLO+NLL matched corrections affect the transverse-momentum distributions. 
We find that the NLO+NLL corrections can change the shape of the $p_T$ distribution 
considerably and thus generally cannot be taken into account by using a simple 
$K$-factor. 

The NLO+NLL matched cross sections and $p_T$ distributions presented in 
this paper constitute the state-of-the-art QCD predictions for stop and sbottom 
production and can be employed to improve current and future searches at the 
Tevatron and LHC.

\vfill

\section*{Acknowledgments}
We would like to thank Sven Moch, James Stirling, Robert Thorne and 
Andreas Vogt for valuable discussions and the MSTW collaboration for providing
us with parton distribution functions with improved accuracy in the evolution 
at large $x$. This work has been supported in part by the Helmholtz Alliance
``Physics at the Terascale'', the DFG Graduiertenkolleg ``Elementary
Particle Physics at the TeV Scale'', the Foundation for Fundamental
Research of Matter (FOM), the National Organization for Scientific
Research (NWO), the DFG SFB/TR9 ``Computational Particle Physics'',
and the European Community's Marie-Curie Research Training Network
under contract MRTN-CT-2006-035505 ``Tools and Precision Calculations
for Physics Discoveries at Colliders''. SB would like to thank the Nikhef 
Theory Group, and MK would like to thank the Institute for High Energy Physics (IFAE) 
at the Universitat Aut\'{o}noma de
Barcelona and the CERN TH unit for their hospitality. AK
acknowledges the hospitality of the Institute of Theoretical Physics at the University of
Warsaw. 

\clearpage

\FIGURE{
\hspace{-1mm}\epsfig{file=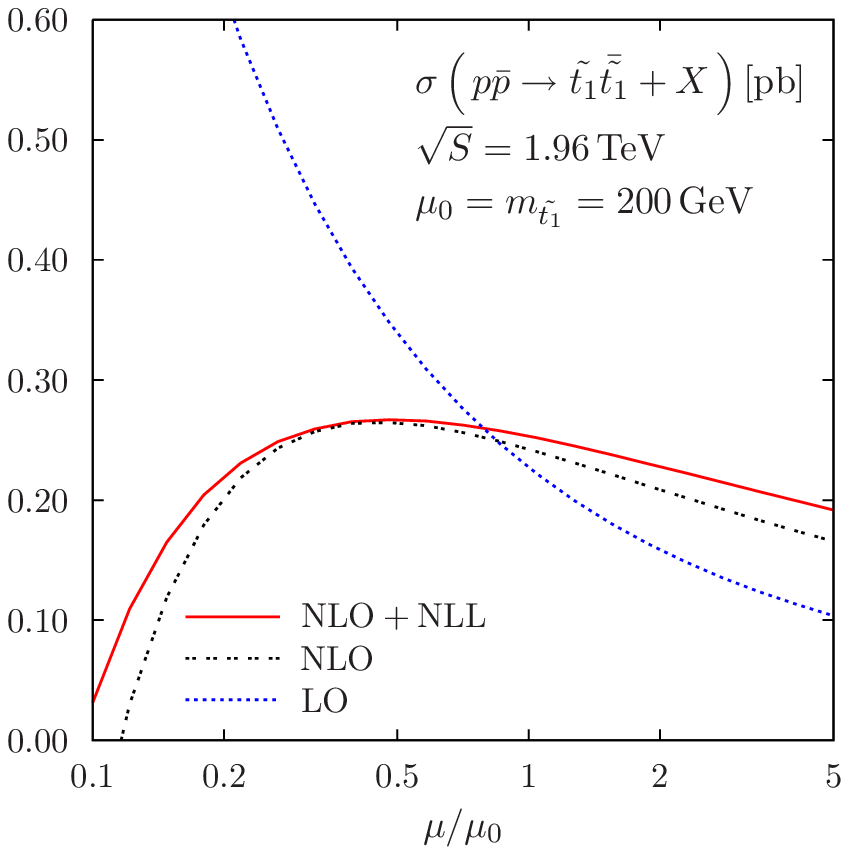, width=0.45\columnwidth}
\hspace*{4mm}\epsfig{file=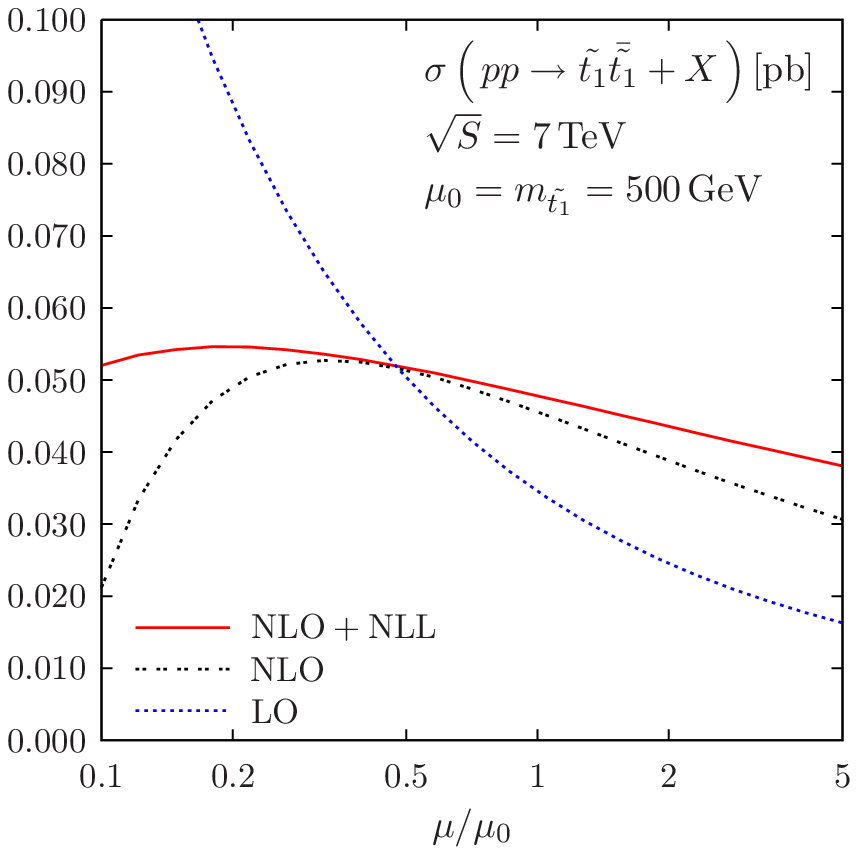, width=0.45\columnwidth}\\[5mm]
\hspace*{2mm}\epsfig{file=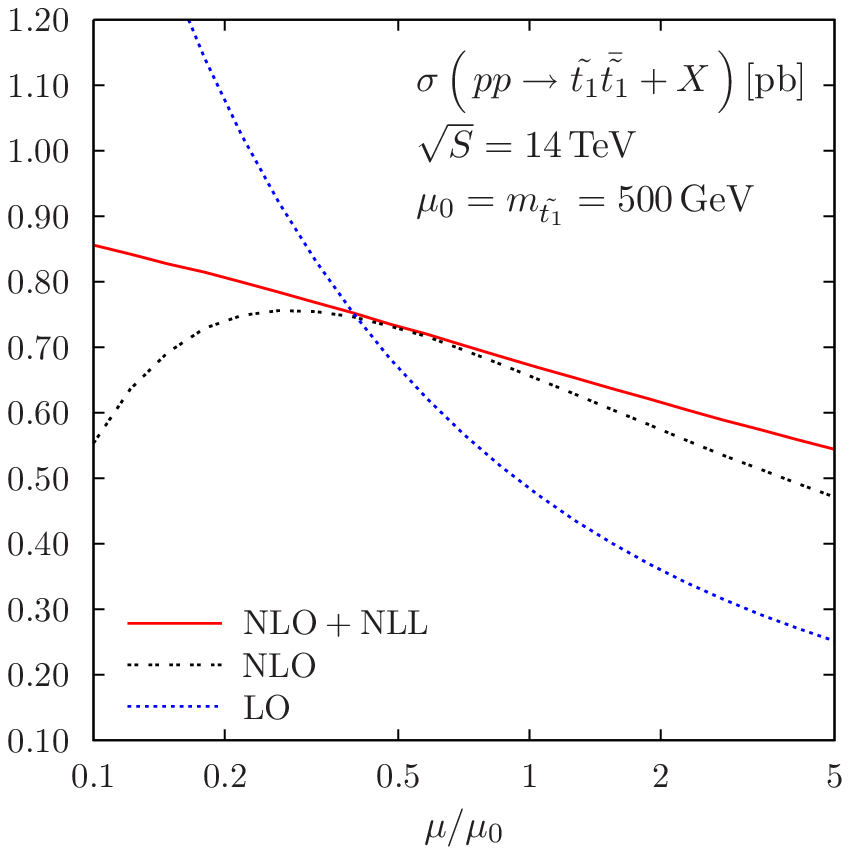, width=0.45\columnwidth}
\caption{Scale dependence of the LO, NLO and NLO+NLL cross section for 
         stop-antistop production at the Tevatron and the LHC.}
\label{fig:scale}
 }

\FIGURE{
\begin{tabular}{ll}
\hspace*{-1mm}\epsfig{file=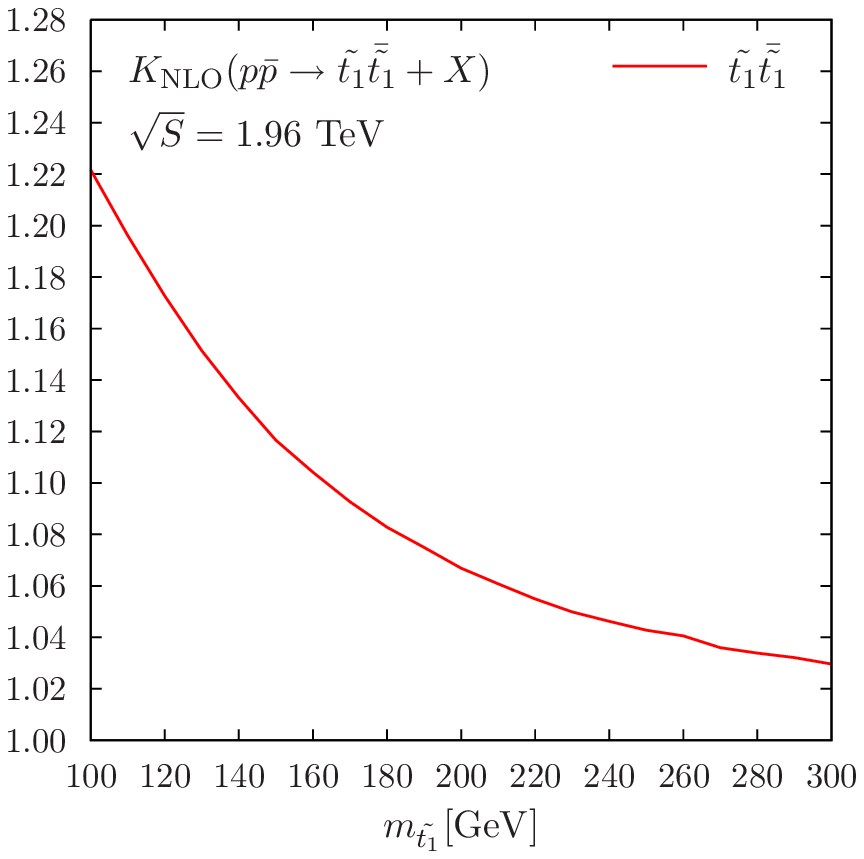, width=0.45\columnwidth}
\hspace*{4mm}\epsfig{file=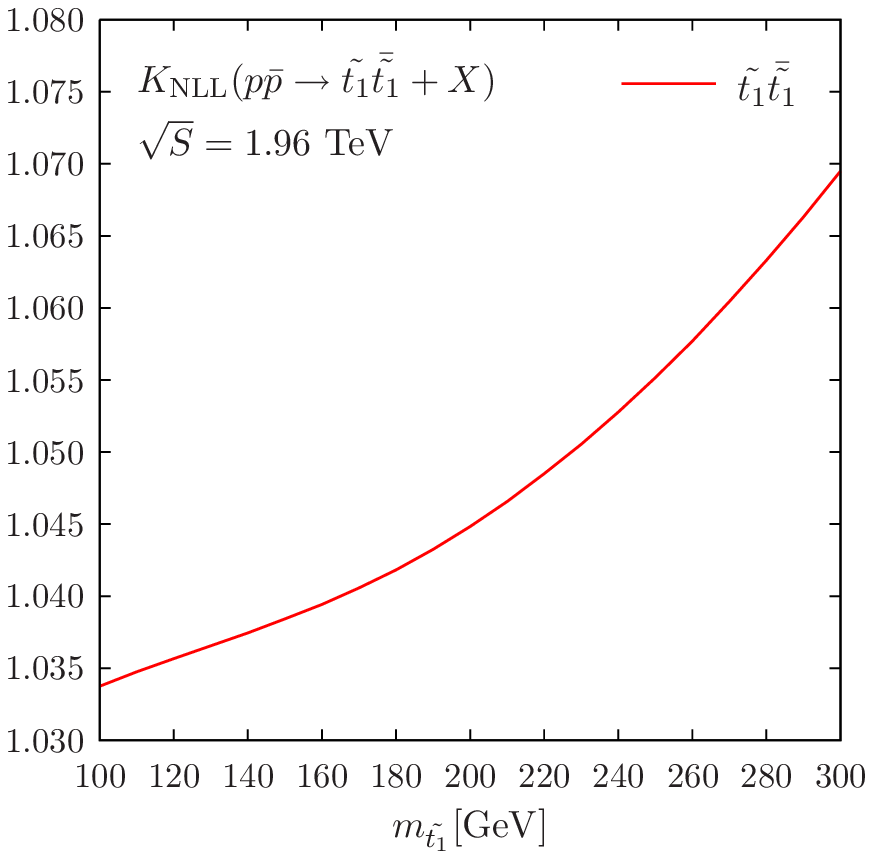, width=0.45\columnwidth}\\[2mm]
\hspace*{-1mm}\epsfig{file=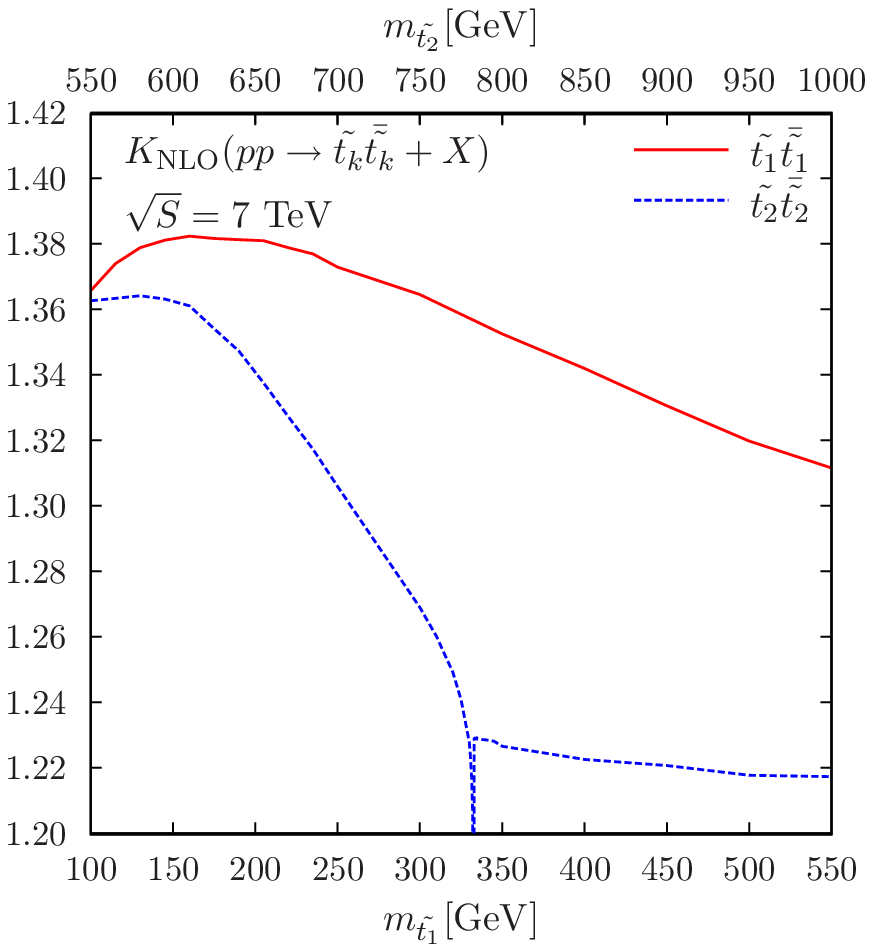, width=0.45\columnwidth}
\hspace*{4mm}\epsfig{file=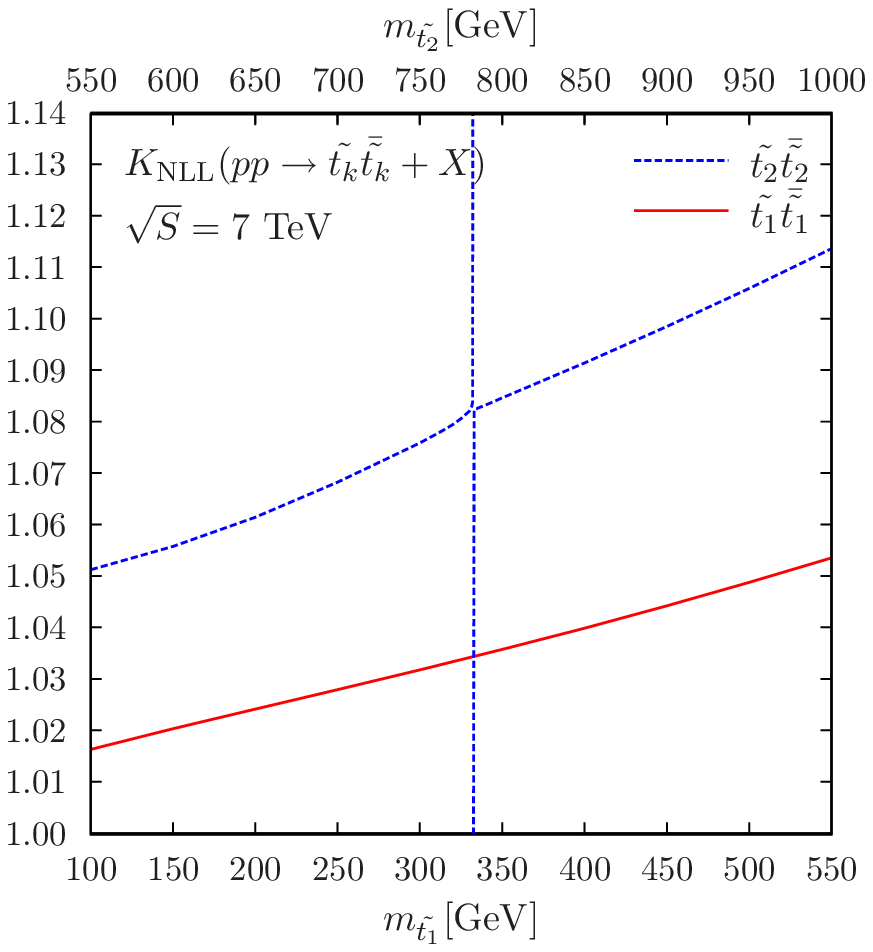, width=0.45\columnwidth}\\[2mm]
\hspace*{-1mm}\epsfig{file=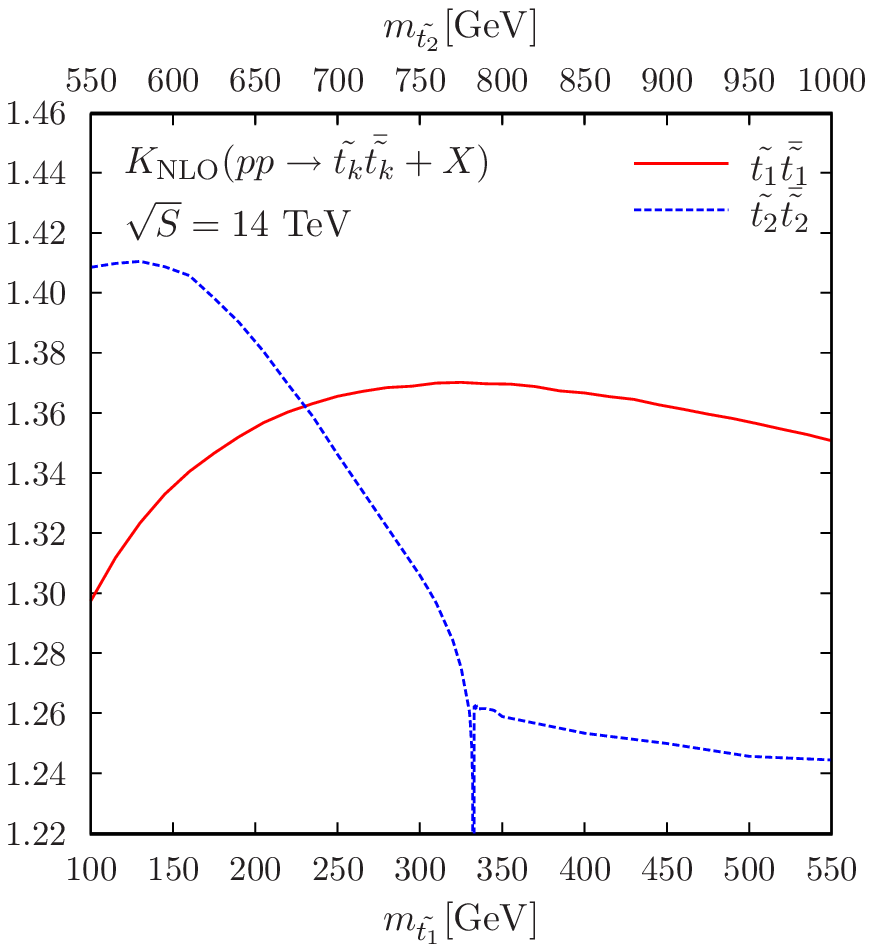, width=0.45\columnwidth}
\hspace*{4mm}\epsfig{file=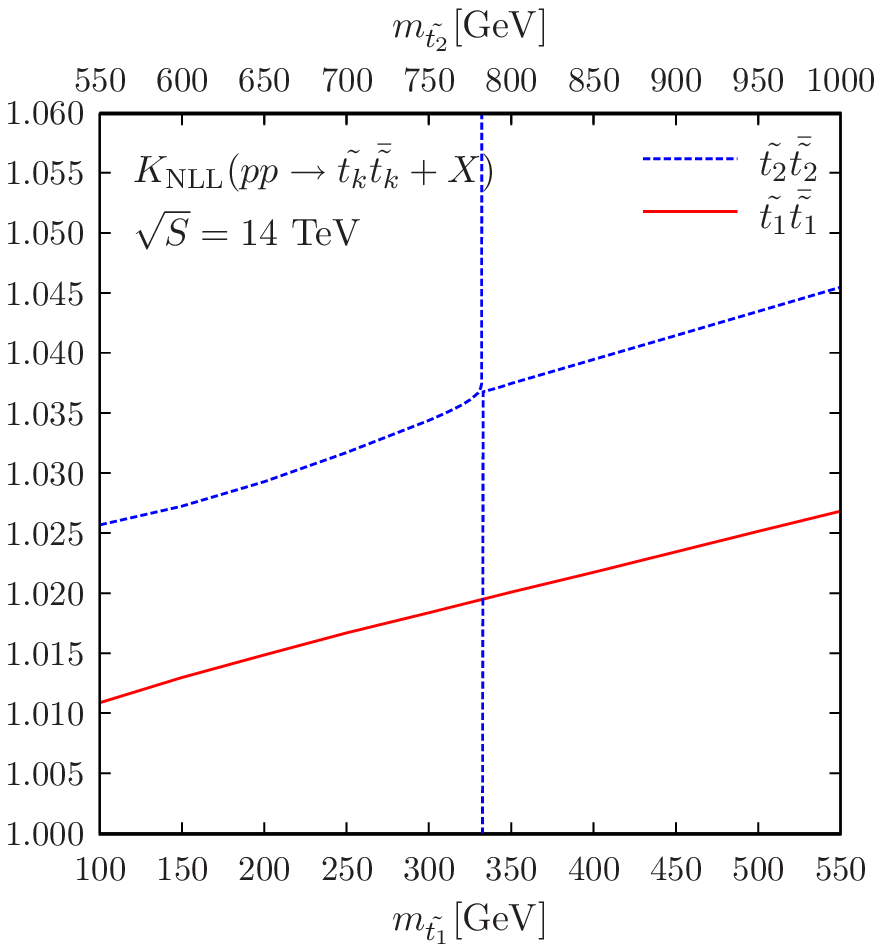, width=0.45\columnwidth}
\end{tabular}
\caption{The NLO and NLL $K$-factors for stop-antistop production at the 
         Tevatron and the LHC as a function of the stop mass. The scale has 
         been set to the stop mass, i.e.~$\mu=m_{\tilde{t}_k}$.}
\label{fig:K_NLO_K_NLL}
}

\FIGURE{
\hspace{-1mm}\epsfig{file=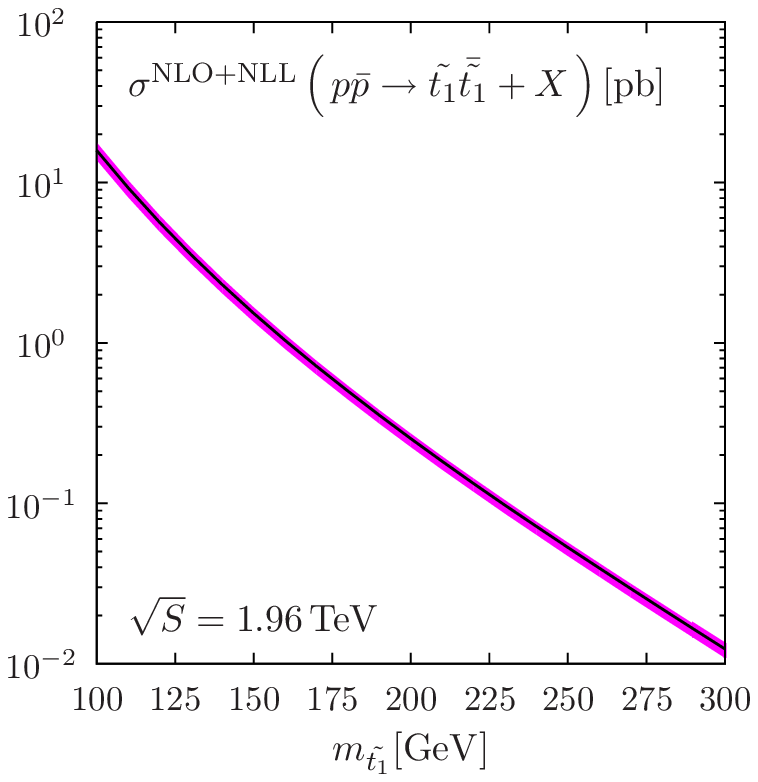, width=0.45\columnwidth}
\hspace{4mm}\epsfig{file=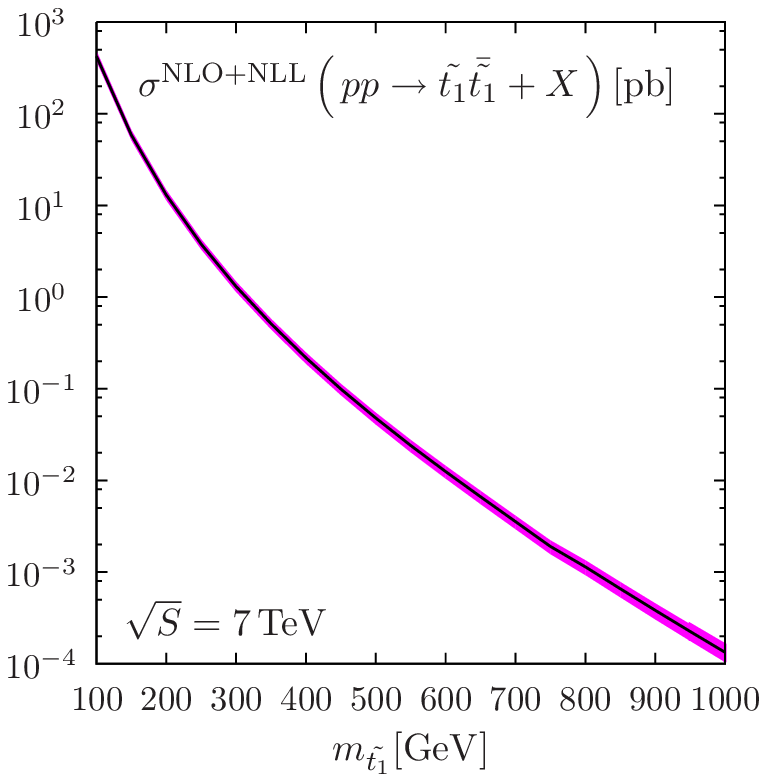, width=0.45\columnwidth}\\[5mm]
\hspace{2mm}\epsfig{file=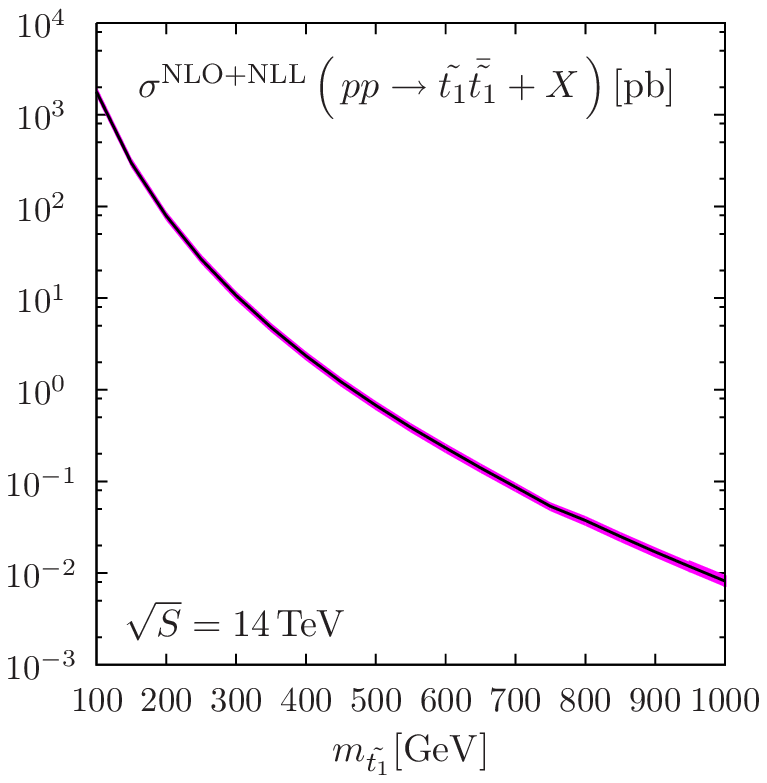, width=0.45\columnwidth}
\caption{Total NLO+NLL stop-pair cross section at the Tevatron and the LHC as 
         a function of the stop mass. The error band corresponds to the scale 
         and pdf uncertainty of the prediction, added in quadrature.}
\label{fig:total}
 }

\clearpage

\renewcommand{\arraystretch}{1.5}
\TABLE{
\begin{footnotesize}
 \begin{tabular}{c|c|c|c||c|c|c }
 \multicolumn{7}{c}{\normalsize $p\bar{p} \to \tilde{t_1}\bar{\tilde{t_1}}$ at $\sqrt{S}=1.96\TeV$}
 \\[1mm] \hline 
 \multicolumn{1}{c}{} & \multicolumn{3}{c||}{MSTW2008} & \multicolumn{3}{c}{CTEQ6.6M} \\[1mm] \hline
$m_{\tilde{t_1}} \; [\mathrm{GeV}]$ & 100 & 200 & 300 & 100 & 200 & 300   \\  \hline \hline
$(\sigma\pm\Delta\sigma_{\mu})_\mathrm{LO} \; [\mathrm{pb}]$ & $12.6 ^{+6.4}_{-3.9}$ & $0.227 ^{+0.112}_{-0.068}$ & $(1.12 ^{+0.57}_{-0.35})\times 10^{-2}$ & $10.3 ^{+4.3}_{-2.9}$ & $0.210 ^{+0.091}_{-0.059}$ & $(1.20 ^{+0.51}_{-0.33})\times 10^{-2}$\\  \hline 
$(\sigma\pm\Delta\sigma_{\mu})_\mathrm{NLO} \; [\mathrm{pb}]$ & $15.3 ^{+2.0}_{-2.4}$ & $0.242 ^{+0.022}_{-0.034}$ & $(1.15 ^{+0.12}_{-0.17})\times 10^{-2}$ & $14.7 ^{+1.8}_{-2.2}$ & $0.249 ^{+0.022}_{-0.034}$ & $(1.23 ^{+0.13}_{-0.18})\times 10^{-2}$\\  \hline
$(\sigma\pm\Delta\sigma_{\mu})_\mathrm{NLO+NLL} \; [\mathrm{pb}]$ & $15.9 ^{+1.6}_{-1.8}$ & $0.253 ^{+0.014}_{-0.025}$ & $(1.24 ^{+0.07}_{-0.13})\times 10^{-2}$ & $15.1 ^{+1.4}_{-1.6}$ & $0.260 ^{+0.014}_{-0.025}$ & $(1.31 ^{+0.08}_{-0.14})\times 10^{-2}$\\  \hline
$\Delta\mathrm{pdf}_\mathrm{NLO} \; [\%]$ & $\pm 6.6$ & $\pm 5.3$ & $\pm 5.3$ & $\pm 11$ & $\pm 11$ & $\pm 11$\\  \hline 
$\mathrm{K}_{\mathrm{NLO}}$ & 1.22 & 1.07 & 1.03 & 1.43 & 1.19 & 1.10\\ \hline 
$\mathrm{K}_{\mathrm{NLL}}$ & 1.03 & 1.05 & 1.07 & 1.03 & 1.04 & 1.07\\
 \end{tabular}

\caption{The LO, NLO and NLO+NLL cross sections for stop-antistop production 
         at the Tevatron ($\sqrt{S}$=1.96 TeV), including errors due to scale 
         variation ($\Delta\sigma_{\mu}$) in the range 
         $m_{\tilde{t}_1}/2 \le \mu \le 2m_{\tilde{t}_1}$. Results are shown 
         for two pdf parametrizations (MSTW08 and CTEQ6) with the corresponding
         90\% C.L. pdf error estimates.}
\label{tab:tevatron}
\end{footnotesize}
}

\TABLE{
\begin{small}
 \begin{tabular}{c|c|c||c|c}
 \multicolumn{5}{c}{\normalsize $p{p} \to \tilde{t_1}\bar{\tilde{t_1}}$ at $\sqrt{S}=7\TeV$}
 \\[1mm]  \hline 
 \multicolumn{1}{c}{} & \multicolumn{2}{c||}{MSTW2008} & \multicolumn{2}{c}{CTEQ6.6M} \\[1mm] \hline
$m_{\tilde{t_1}} \; [\mathrm{GeV}]$ & 100 & 400 & 100 & 400  \\  \hline \hline
$(\sigma\pm\Delta\sigma_{\mu})_\mathrm{LO} \; [\mathrm{pb}]$ & $305 ^{+114}_{-77}$ & $0.156 ^{+0.070}_{-0.044}$ & $265 ^{+95}_{-65}$ & $0.119 ^{+0.048}_{-0.032}$\\  \hline 
$(\sigma\pm\Delta\sigma_{\mu})_\mathrm{NLO} \; [\mathrm{pb}]$ & $416^{+64}_{-59}$ & $0.209 ^{+0.027}_{-0.031}$ & $384 ^{+57}_{-52}$ & $0.202 ^{+0.025}_{-0.028}$\\  \hline 
$(\sigma\pm\Delta\sigma_{\mu})_\mathrm{NLO+NLL} \; [\mathrm{pb}]$ & $423^{+60}_{-46}$ & $0.218 ^{+0.020}_{-0.020}$ & $390 ^{+53}_{-41}$ & $0.209 ^{+0.018}_{-0.019}$\\  \hline
$\Delta\mathrm{pdf}_\mathrm{NLO} \; [\%]$ & $\pm 3.9$ & $\pm 10$ & $\pm 3.4$ & $\pm 17$\\  \hline 
$\mathrm{K}_{\mathrm{NLO}}$ & 1.37 & 1.34 & 1.45 & 1.70  \\ \hline 
$\mathrm{K}_{\mathrm{NLL}}$ & 1.02 & 1.04 & 1.02 & 1.04 \\
 \end{tabular}
\\[5mm]
 \begin{tabular}{c|c|c||c|c}
 \multicolumn{5}{c}{\normalsize $p{p} \to \tilde{t_2}\bar{\tilde{t_2}}$ at $\sqrt{S}=7\TeV$}
 \\[1mm]  \hline 
 \multicolumn{1}{c}{} & \multicolumn{2}{c||}{MSTW2008} & \multicolumn{2}{c}{CTEQ6.6M} \\[1mm] \hline
$m_{\tilde{t_2}} \; [\mathrm{GeV}]$ & 600 & 1000 & 600 & 1000  \\  \hline \hline
$(\sigma\pm\Delta\sigma_{\mu})_\mathrm{LO} \; [\mathrm{pb}]$ & $(9.06 ^{+4.22}_{-2.66})\times 10^{-3}$ & $(9.64 ^{+4.83}_{-2.97})\times 10^{-5}$ & $(6.63 ^{+2.70}_{-1.78})\times 10^{-3}$ & $(6.76 ^{+2.86}_{-1.88})\times 10^{-5}$\\  \hline 
$(\sigma\pm\Delta\sigma_{\mu})_\mathrm{NLO} \; [\mathrm{pb}]$ & $(1.23 ^{+0.18}_{-0.20})\times 10^{-2}$ & $(1.17 ^{+0.18}_{-0.20})\times 10^{-4}$ & $(1.27 ^{+0.18}_{-0.20})\times 10^{-2}$ & $(1.50 ^{+0.18}_{-0.24})\times 10^{-4}$\\  \hline 
$(\sigma\pm\Delta\sigma_{\mu})_\mathrm{NLO+NLL} \; [\mathrm{pb}]$ & $(1.30 ^{+0.13}_{-0.12})\times 10^{-2}$ & $(1.31 ^{+0.05}_{-0.09})\times 10^{-4}$ & $(1.33 ^{+0.13}_{-0.13})\times 10^{-2}$ & $(1.64 ^{+0.07}_{-0.11})\times 10^{-4}$\\  \hline
$\Delta\mathrm{pdf}_\mathrm{NLO} \; [\%]$ & $\pm 15$ & $\pm 23$ & $\pm 26$ & $\pm 46$\\  \hline 
$\mathrm{K}_{\mathrm{NLO}}$ & 1.36 & 1.22 & 1.92 & 2.21  \\ \hline 
$\mathrm{K}_{\mathrm{NLL}}$ & 1.06 & 1.11 & 1.05 & 1.10 \\
 \end{tabular}

\caption{The LO, NLO and NLO+NLL cross sections for stop-antistop production
         at the LHC ($\sqrt{S}$=7 TeV), including errors due to scale variation
         ($\Delta\sigma_{\mu}$) in the range 
         $m_{\tilde{t}}/2 \le \mu \le 2m_{\tilde{t}}$. Results are shown for 
         two pdf parametrizations (MSTW08 and CTEQ6) with the corresponding 
         90\% C.L. pdf error estimates.}
\label{tab:lhc7}
\end{small}
}

\TABLE{
\begin{small}
 \begin{tabular}{c|c|c||c|c}
 \multicolumn{5}{c}{\normalsize $p{p} \to \tilde{t_1}\bar{\tilde{t_1}}$ at $\sqrt{S}=14\TeV$}
 \\[1mm]\hline 
 \multicolumn{1}{c}{} & \multicolumn{2}{c||}{MSTW2008} & \multicolumn{2}{c}{CTEQ6.6M} \\[1mm] \hline
$m_{\tilde{t_1}} \; [\mathrm{GeV}]$ & 100 & 400 & 100 & 400  \\  \hline \hline
$(\sigma\pm\Delta\sigma_{\mu})_\mathrm{LO} \; [\mathrm{pb}]$ & $(1.35 ^{+0.41}_{-0.29})\times 10^{3}$ & $1.67 ^{+0.62}_{-0.42} $& $(1.22 ^{+0.35}_{-0.26})\times 10^{3}$ & $1.40 ^{+0.49}_{-0.34}$\\  \hline 
$(\sigma\pm\Delta\sigma_{\mu})_\mathrm{NLO} \; [\mathrm{pb}]$ & $(1.75 ^{+0.26}_{-0.22})\times 10^{3}$ & $2.29 ^{+0.25}_{-0.29}$ & $(1.63 ^{+0.23}_{-0.19})\times 10^{3}$ & $2.14^{+0.24}_{-0.26}$\\  \hline 
$(\sigma\pm\Delta\sigma_{\mu})_\mathrm{NLO+NLL} \; [\mathrm{pb}]$ & $(1.77 ^{+0.24}_{-0.17})\times 10^{3}$ & $2.34 ^{+0.21}_{-0.21}$ & $(1.65 ^{+0.22}_{-0.16})\times 10^{3}$ & $2.19 ^{+0.20}_{-0.19}$\\  \hline
$\Delta\mathrm{pdf}_\mathrm{NLO} \; [\%]$ & $\pm 2.8$ & $\pm 6.2$ & $\pm 2.6$ & $\pm 8.6$\\  \hline 
$\mathrm{K}_{\mathrm{NLO}}$ & 1.30 & 1.37 & 1.34 & 1.53  \\ \hline 
$\mathrm{K}_{\mathrm{NLL}}$ & 1.01 & 1.02 & 1.01 & 1.02 \\
 \end{tabular}
\\[5mm]
 \begin{tabular}{c|c|c||c|c}
 \multicolumn{5}{c}{\normalsize $p{p} \to \tilde{t_2}\bar{\tilde{t_2}}$ at $\sqrt{S}=14\TeV$}
 \\[1mm]  \hline
 \multicolumn{1}{c}{} & \multicolumn{2}{c||}{MSTW2008} & \multicolumn{2}{c}{CTEQ6.6M} \\[1mm] \hline
$m_{\tilde{t_2}} \; [\mathrm{GeV}]$ & 600 & 1000 & 600 & 1000  \\  \hline \hline
$(\sigma\pm\Delta\sigma_{\mu})_\mathrm{LO} \; [\mathrm{pb}]$ & $0.167 ^{+0.065}_{-0.043}$ & $(6.13 ^{+2.51}_{-1.65})\times 10^{-3}$ & $0.135^{+0.048}_{-0.033}$ & $(4.71 ^{+1.72}_{-1.17})\times 10^{-3}$\\  \hline 
$(\sigma\pm\Delta\sigma_{\mu})_\mathrm{NLO} \; [\mathrm{pb}]$ & $0.235 ^{+0.030}_{-0.031}$ &$(7.63 ^{+0.65}_{-0.92})\times 10^{-3}$ & $0.225 ^{+0.027}_{-0.029}$ &  {$(7.65 ^{+0.62}_{-0.90})\times 10^{-3}$}\\  \hline 
$(\sigma\pm\Delta\sigma_{\mu})_\mathrm{NLO+NLL} \; [\mathrm{pb}]$ & $0.242 ^{+0.024}_{-0.022}$ & $(7.98 ^{+0.36}_{-0.51})\times 10^{-3}$ & $0.230^{+0.022}_{-0.020}$ & $(7.97 ^{+0.35}_{-0.50})\times 10^{-3}$\\  \hline
$\Delta\mathrm{pdf}_\mathrm{NLO} \; [\%]$ & $\pm 8.3$ & $\pm 12$ & $\pm 13$ & $\pm 21$\\  \hline 
$\mathrm{K}_{\mathrm{NLO}}$ & 1.41 & 1.24 & 1.66 & 1.62  \\ \hline 
$\mathrm{K}_{\mathrm{NLL}}$ & 1.03 & 1.05 & 1.03 & 1.04 \\
 \end{tabular}

\caption{The LO, NLO and NLO+NLL cross sections for stop-antistop production 
         at the LHC ($\sqrt{S}$=14 TeV), including errors due to scale 
         variation ($\Delta\sigma_{\mu}$) in the range 
         $m_{\tilde{t}}/2 \le \mu \le 2m_{\tilde{t}}$. Results are shown for 
         two pdf parametrizations (MSTW08 and CTEQ6) with the corresponding 
         90\% C.L. pdf error estimates.}
\label{tab:lhc14}
\end{small}
}

\clearpage

 \FIGURE{
\hspace{-1mm}\epsfig{file=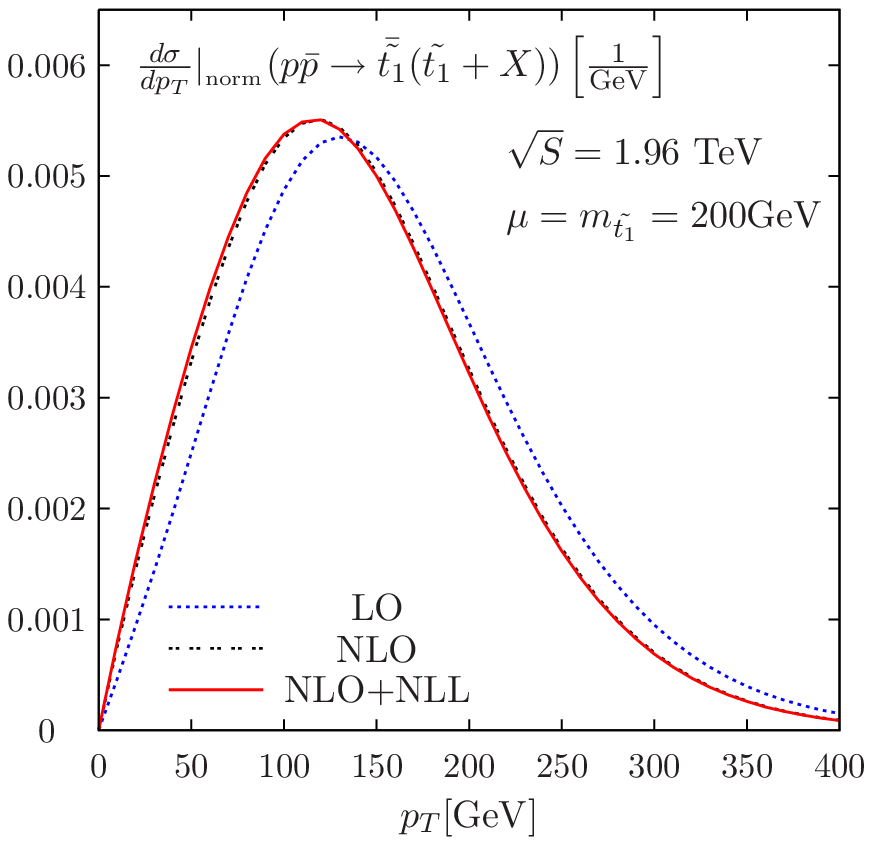, width=0.45\columnwidth}
\hspace{4mm}\epsfig{file=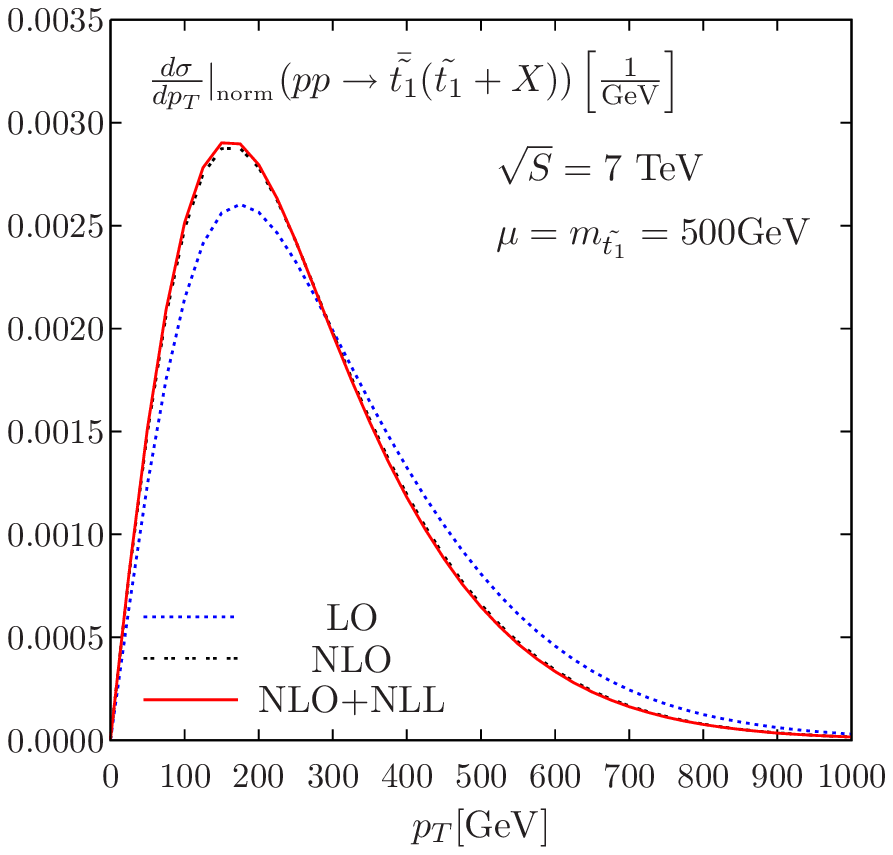, width=0.46\columnwidth}\\[5mm]
\hspace{2mm}\epsfig{file=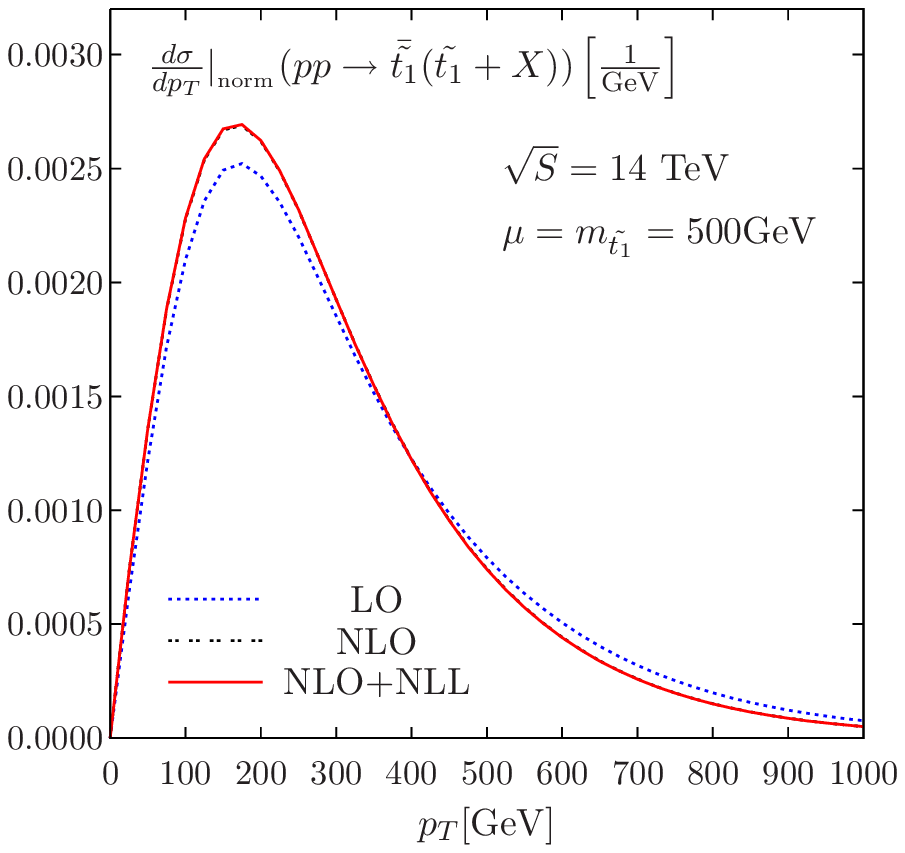, width=0.45\columnwidth}
\caption{Normalized LO, NLO and NLO+NLL transverse-momentum distributions for 
         stop-antistop production at the Tevatron and the LHC for 
         $\mu=m_{\tilde{t}_1}$.}
\label{fig:pt_m}
 }

\FIGURE{
\hspace*{-1mm}\epsfig{file=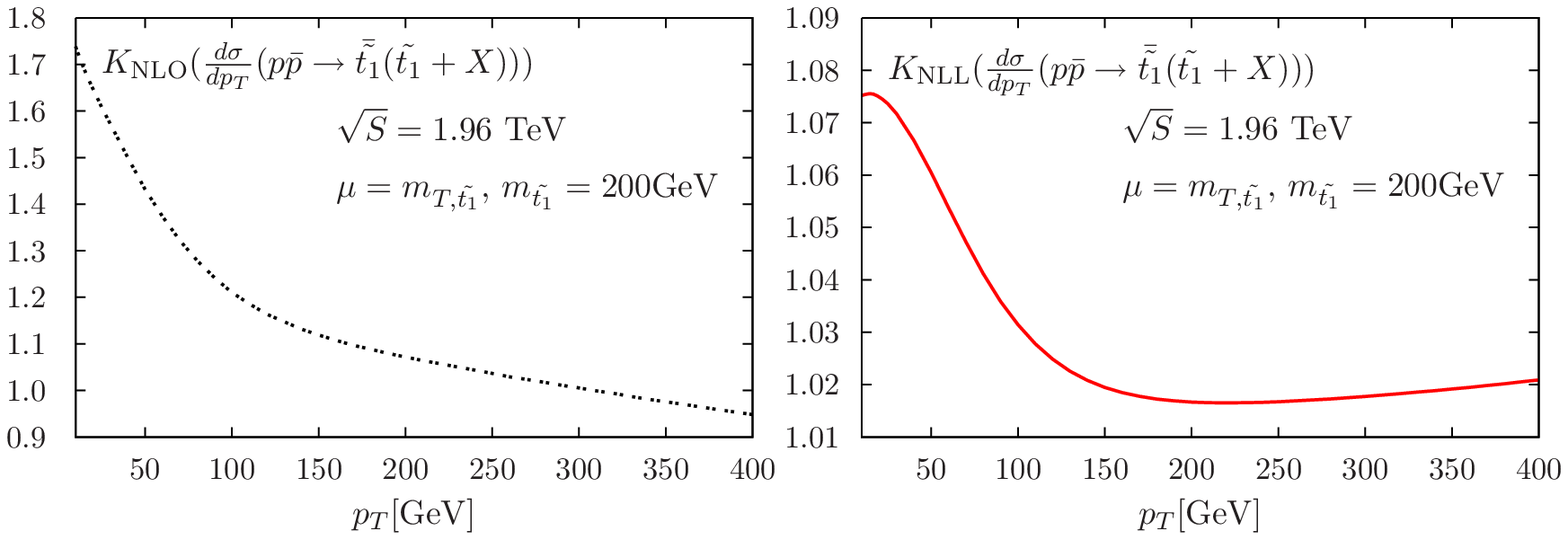, width=0.9\columnwidth}\\[5mm]
\epsfig{file=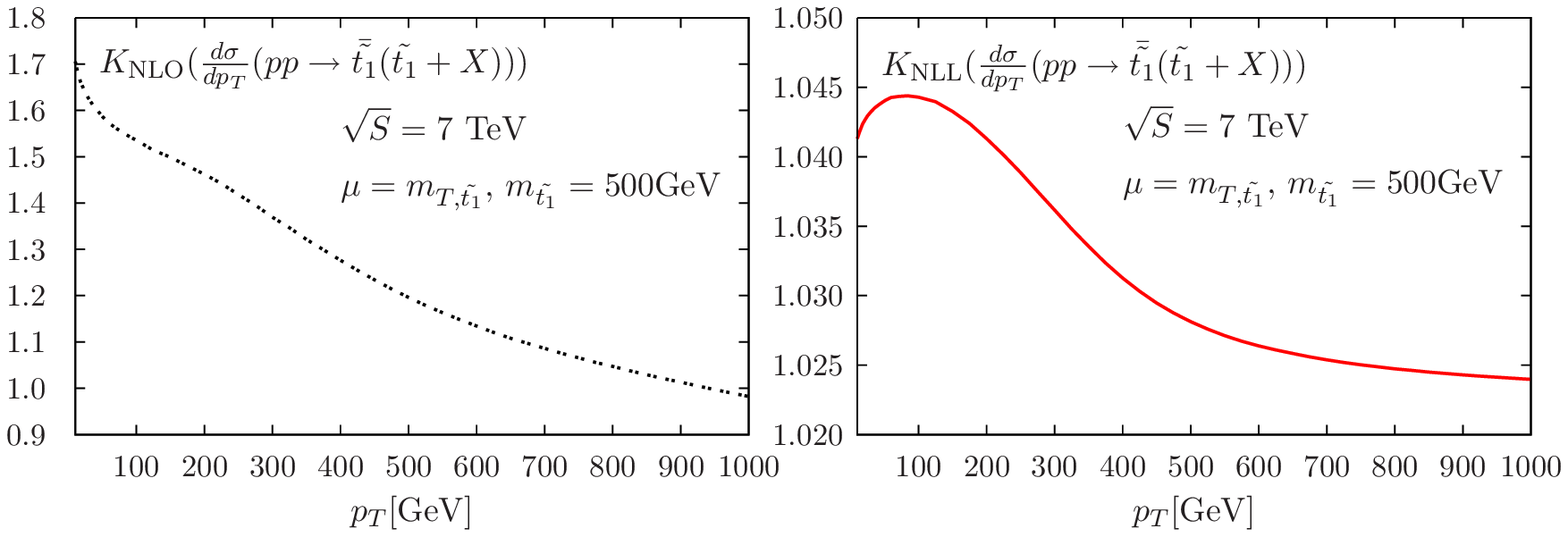, width=0.9\columnwidth}\\[5mm]
\epsfig{file=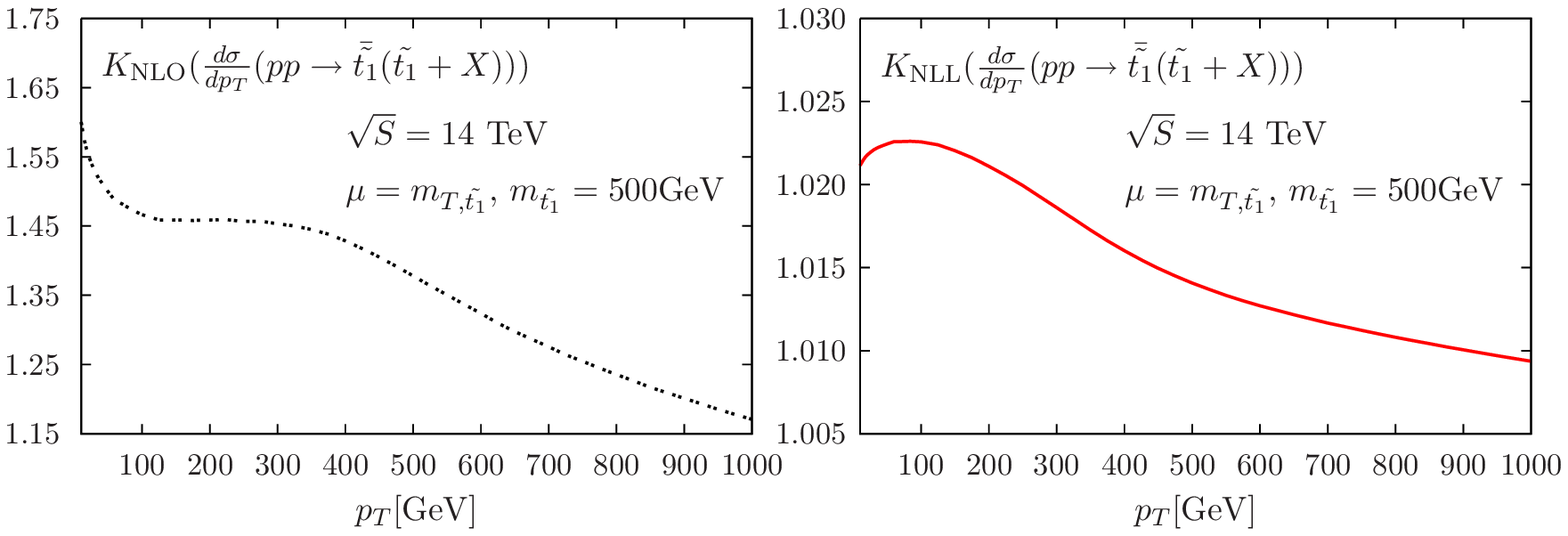, width=0.9\columnwidth}
\caption{Transverse-momentum dependence of the NLO and NLL $K$-factors for 
         stop-antistop production at the Tevatron and the LHC for 
         $\mu=m_{T,\tilde{t}_1}$.} 
\label{fig:pt_mt}
 }

\clearpage

\begin{appendix}
\section{SUSY parameter dependence of stop and sbottom cross sections}

In this appendix we shall investigate the dependence of the NLO+NLL stop and 
sbottom cross-section predictions on the supersymmetric model parameters that
enter beyond LO, i.e.~the mixing angle and the masses of the light-flavour 
squarks and the gluino. In the case of sbottom production we shall in
addition quantify the impact of the bottom-quark-induced reaction channels 
$b\bar{b}\to \tilde{b}_k\bar{\tilde{b}}_k\,$ and 
$\,bb/\bar{b}\bar{b}\to \tilde{b}_k\tilde{b}_k/\bar{\tilde{b}}_k\bar{\tilde{b}}_k$, 
involving contributions from $t$-channel gluino exchange.
It will be demonstrated that the contributions of these partonic reaction 
channels are strongly suppressed numerically. Thus, for all practical purposes,
cross-section predictions obtained for stop-pair production also apply to 
sbottom-pair production when the same input parameters are adopted.

As in the main body of the paper we choose the SPS1a' benchmark 
scenario~\cite{AguilarSaavedra:2005pw} as our default. The SPS1a' masses and 
mixings relevant for stop and sbottom hadroproduction are collected in 
Table~\ref{tab:sps1a}.
\TABLE{
\begin{small}
 \begin{tabular}{|c|c||c|c||c|c|}
\multicolumn{6}{c}{SPS1a'}\\\hline
$m_{\tilde{t}_1}$ & 367~GeV & $m_{\tilde{b}_1}$ & 509~GeV & $m_{\tilde{q}}$ & 560~GeV \\ 
$m_{\tilde{t}_2}$ & 590~GeV & $m_{\tilde{b}_2}$ & 549~GeV & $m_{\tilde{g}}$ & 610~GeV \\
$\sin2\theta_{\tilde{t}}$ & 0.932 & $\sin2\theta_{\tilde{b}}$ & 0.652 & & \\\hline
\end{tabular}
\caption{Masses and mixings for the SPS1a' benchmark 
         scenario~\cite{AguilarSaavedra:2005pw} obtained using 
         {\tt SPheno}~\cite{Porod:2003um} with the Standard Model input
         parameters $m_t=172.5$ GeV and $\alpha_{\rm s}(M_{Z}) = 0.120$.}
\label{tab:sps1a}
\end{small}}
Note that the stop and sbottom masses predicted by the SPS1a' and other 
commonly used benchmark scenarios are beyond the reach of the Tevatron 
searches, as the corresponding production cross sections are too small. 
The SPS1a' NLO+NLL cross sections for stop and sbottom production at
the LHC are collected in Table~\ref{tab:spsxs}. 
\TABLE{
\begin{small}
\begin{tabular}{|c|c|c|}
\hline
& \multicolumn{2}{c|}{$\sigma_{\rm NLO+NLL}$ [pb]} \\
\cline{2-3}
\raisebox{3.75mm}[4mm][2mm]{SPS1a'} & LHC @ 7~TeV & LHC @ 14~TeV \\\hline
$\tilde{t}_1\bar{\tilde{t}}_1$ & 0.379 & 3.71  \\
$\tilde{t}_2\bar{\tilde{t}}_2$ & $1.48\times 10^{-2}$ & 0.268 \\
$\tilde{b}_1\bar{\tilde{b}}_1$ & $4.23 \times 10^{-2}$ & 0.611  \\
$\tilde{b}_2\bar{\tilde{b}}_2$ & $2.51 \times 10^{-2}$ & 0.405 \\\hline
\end{tabular}
\caption{NLO+NLL SUSY-QCD cross sections for stop and sbottom pair production 
         at the LHC for the SPS1a' benchmark scenario. The MSTW pdfs 
         have been adopted and the scale has been set to the mass of the 
         particles produced.}
\label{tab:spsxs}
\end{small}}
\noindent
From the cross-section predictions one can conclude that only the lighter of
the SPS1a' stop mass eigenstates might be detected during the initial phase of 
LHC data taking at 7~TeV with 1~fb$^{-1}$ of integrated~lumi\-no\-si\-ty. 
Also a dedicated search for sbottom production in SPS1a'-like scenarios will 
on\-ly be possible with higher LHC energies.

Therefore, to address the SUSY-para\-meter dependence and to study the impact 
of bottom-quark-induced sbottom-pair pro\-duction, we consider two 
different sce\-na\-rios that are within the reach of the Tevatron and
the early LHC phase. As we did in the main body of the paper, we use stop and 
sbottom masses of $100\,(200)$~GeV at the Tevatron and $100\,(400)$~GeV at the LHC, 
respectively, and present results for various choices of the mixing angle and the 
light-flavour squark and gluino masses, see Table~\ref{tab:susydep}. Note that the 
NLL resummation only depends on the final-state particle mass. The dependence 
on the other SUSY parameters enters exclusively through the NLO virtual corrections. 
The numbers listed in Table~\ref{tab:susydep} reveal that the dependence of the 
cross section on the mixing angle, the gluino mass, and the light-flavour squark 
masses is small indeed, with variations of at most 2\% both at the Tevatron and at 
the LHC. 
\TABLE{
\begin{small}
 \begin{tabular}{r|c|c||c|c||c|c}
  \multicolumn{7}{c}{\normalsize $\sigma(pp/p\bar{p} \to
    \tilde{t}_1\bar{\tilde{t}}_1)$~[pb]}
 \\[1mm]  \cline{1-7} 
& \multicolumn{2}{c||}{Tevatron (1.96 TeV)}  &
\multicolumn{2}{c||}{LHC @ 7~TeV} & \multicolumn{2}{c}{LHC @ 14~TeV}\\
\cline{1-7}
 $m_{\tilde{t}_1}$~[GeV]      & 100   &    200 & 100 & 400 &100 & 400 \\\hline
SPS1a' default                  & 15.9  & 0.253 & 423 & 0.218 & $1.77 \times 10^{3}$ & 2.34 \\\hline
$\sin{2{\theta}_{\tilde{t}}} = -1$      & 15.9  & 0.255 &  425 & 0.222 & $1.78 \times 10^{3}$  & 2.39 \\
                                0      & 15.9  & 0.254 &  423 & 0.219 & $1.77 \times 10^{3}$  & 2.36 \\
                              +1     & 15.9  & 0.253 &  423 & 0.218 & $1.77 \times 10^{3}$  & 2.33 \\\hline
$m_{\tilde{q}} = 200$~GeV & 15.8  & 0.248 & 423 & 0.217 & $1.77 \times 10^{3}$  & 2.34 \\
                       500~GeV   &15.9   & 0.252 & 423 & 0.218 & $1.77 \times 10^{3}$  & 2.34 \\
                      1000~GeV  & 15.9  & 0.255 & 423 & 0.219 & $1.77 \times 10^{3}$ & 2.34 \\\hline
$m_{\tilde{g}} = 200$~GeV  & 15.8  & 0.251 & 421 & 0.214 & $1.76 \times 10^{3}$  & 2.29 \\
                        500~GeV  & 15.9  & 0.253 & 423 & 0.217 & $1.77 \times 10^{3}$  & 2.33 \\
                      1000~GeV  & 15.9  & 0.254 & 424 & 0.219 & $1.77 \times 10^{3}$  & 2.34 \\\hline
\end{tabular}

\caption{The NLO+NLL cross sections for stop-antistop production at the 
         Tevatron and the LHC. We compare the SPS1a' default input for the stop
         mixing angle and the light-flavour squark and gluino masses with 
         various other choices for these SUSY parameters. Note that only one 
         parameter is changed at a time, while the others are kept at their  
         default values. The MSTW pdfs have been adopted and the scale has
         been set to the final-state stop mass.}
\label{tab:susydep}
\end{small}
}

Cross sections for $\tilde{b}_1\bar{\tilde{b}}_1$ production are collected in 
Table~\ref{tab:susydep_sbsbx}. We compare the LO bottom-quark-induced 
contributions with the LO and NLO+NLL predictions based on the stop-like 
contributions that exclude bottom-quark initial states. Using the notation 
introduced in Sect.~2 and $m_-^2\equiv m_{\tilde g}^2-m_{\tilde b_1}^2$, the 
LO bottom-quark-induced contributions read
\begin{align*}
  \sigma^{(0)}_{b\bar b\to\tilde b_1\bar{\tilde b}_1}
  =&\ \frac{\alpha_{\rm s}^2\pi\,C_F}{N_c\,s}\Biggl[\,
      \biggl(\, \frac{m_{\tilde g}^2s\,[1-\cos(4\theta_{\tilde b})]}
                     {8(m_{\tilde g}^2s+m_-^4)}
          \,-\, \frac{s+2m_{\tilde b_1}^2}{3s}
          \,-\, \frac{1+\cos(4\theta_{\tilde b})}{4}
          \,+\, \frac{s+2m_-^2}{2N_c\,s} \,\biggr)\beta \\[2mm]
   &\hphantom{\frac{\alpha_{\rm s}^2\pi\,C_F}{N_c\,s}A}
   +\,\biggl(\, \frac{m_-^4+sm_{\tilde g}^2}{N_c\,s^2}
          \,-\, \frac{(s+2m_-^2)[3+\cos(4\theta_{\tilde b})]}{8s}
      \biggr)\log\Bigl(\,\frac{1-\beta+2m_-^2/s}{1+\beta+2m_-^2/s}\,\Bigr) 
      \,\Biggr]~.
\end{align*}
As these contributions depend on the gluino mass, we give 
results for $m_{\tilde{g}} = 200, 500$~GeV and 1~TeV. From the numbers 
presented in Table~\ref{tab:susydep_sbsbx} it is clear that the
$b\bar b\to\tilde b_1\bar{\tilde b}_1$ channel 
is always strongly suppressed, with cross sections well below 1\% of the
stop-like contributions.

Bottom-quark-induced $t$-channel gluino exchange also leads to 
$\tilde{b}_1\tilde{b}_1$ and $\bar{\tilde{b}}_1\bar{\tilde{b}}_1$ final states.
The LO cross section for these processes is given by
\begin{align*}
  \sigma^{(0)}_{bb\to\tilde b_1\tilde b_1}
  \!=&\ \frac{\alpha_{\rm s}^2\pi\,C_F}{N_c\,s}\Biggl[\,
      \biggl( \frac{1-N_c}{8N_c}\,[1\!-\cos(4\theta_{\tilde b})]
        \,+\, \frac{2m_{\tilde g}^2s-m_-^4+(2m_{\tilde g}^2s+m_-^4)
                    \cos(4\theta_{\tilde b})}
                   {8(m_-^4+m_{\tilde g}^2s)} \,\biggr)\beta \\[2mm]
   &\hspace*{-10ex}
   +\,\biggl(\, \frac{m_-^4(1-\cos(4\theta_{\tilde b}))+4sm_{\tilde g}^2}
                     {4N_c\,s(s+2m_-^2)}
           \,-\, \frac{(s+2m_-^2)(1-\cos(4\theta_{\tilde b}))}{8s}
      \biggr)\log\Bigl(\,\frac{1-\beta+2m_-^2/s}{1+\beta+2m_-^2/s}\,\Bigr) 
      \,\Biggr]~,
\end{align*}
with the identical expression for the charge conjugate process $\bar{b}\bar{b} \to \bar{\tilde b}_1\bar{\tilde b}_1$. The 
corresponding numerical results for $\tilde{b}_1\tilde{b}_1$  production are listed in Table~\ref{tab:susydep_sbsb}. Also the $\tilde{b}_k\tilde{b}_k$ and 
$\bar{\tilde{b}}_k\bar{\tilde{b}}_k$ processes are suppressed by the small 
bottom-quark pdfs and never exceed the per-mille level with respect 
to $\tilde{b}_k\bar{\tilde{b}}_k$ production.

\TABLE{
\begin{small}
 \begin{tabular}{r|c|c||c|c||c|c}
  \multicolumn{7}{c}{\normalsize $\sigma(pp/p\bar{p} \to
    \tilde{b}_1\bar{\tilde{b}}_1)$~[pb]}
 \\[1mm]  \cline{1-7} 
   & \multicolumn{2}{c||}{Tevatron (1.96 TeV)}  &
\multicolumn{2}{c||}{LHC @ 7~TeV} & \multicolumn{2}{c}{LHC @ 14~TeV}\\
\cline{1-7}
 $m_{\tilde{b}_1}$~[GeV] & 100 & 200 & 100 & 400 &100 & 400 \\\hline
SPS1a' default & & & & & & \\[-1mm]
NLO+NLL & 15.9 & 0.253 & 423 & 0.218 & $1.77\times 10^{3}$ &  2.34\\
LO & 12.6 & 0.227 & 305 & 0.156 & $1.35 \times 10^{3}$ & 1.67 \\
LO $b\bar{b}$-channel only & $0.404\times 10^{-2}$ & $0.330 \times
10^{-4}$ & 0.275 & $0.346 \times 10^{-3}$ & 1.40 & $0.564\times 10^{-2}$ \\\hline
LO $b\bar{b}$-channel only& & & & & & \\[-1mm] 
with $m_{\tilde{g}} \!=\!200$~GeV & $0.986\times 10^{-2}$  & $0.870 \times
10^{-4}$ & 0.659  & $0.667 \times 10^{-3}$ & 3.35 & $0.111 \times 10^{-1}$ \\
500~GeV & $0.454\times 10^{-2}$ & $0.399 \times 10^{-4}$ & 0.309 &
$0.408 \times 10^{-3}$ & 1.58 & $0.665 \times 10^{-2}$\\
1000~GeV & $0.335 \times 10^{-2}$ & $0.220 \times 10^{-4}$ & 0.227 &
$0.220 \times 10^{-3}$ & 1.16 & $0.360\times 10^{-2}$ \\
\hline
\end{tabular}
\caption{The LO and NLO+NLL cross sections for sbottom-antisbottom production 
         at the Tevatron and the LHC. We compare the default SPS1a' prediction for
         the stop-like contributions with the LO contributions induced by 
         bottom-quark initial states. The MSTW pdfs have been adopted and the 
         scale has been set to the final-state sbottom mass.}
\label{tab:susydep_sbsbx}
\end{small}
}

\TABLE{
\begin{small}
 \begin{tabular}{r|c|c||c|c||c|c}
  \multicolumn{7}{c}{\normalsize $\sigma(pp/p\bar{p} \to
    \tilde{b}_1\tilde{b}_1)$~[pb]}
 \\[1mm]  \cline{1-7} 
   & \multicolumn{2}{c||}{Tevatron (1.96 TeV)}  &
\multicolumn{2}{c||}{LHC @ 7~TeV} & \multicolumn{2}{c}{LHC @ 14~TeV}\\
\cline{1-7}
 $m_{\tilde{b}_1}$~[GeV] & 100 & 200 & 100 & 400 &100 & 400 \\\hline
SPS1a' default & $0.111\times 10^{-2}$ & $0.188 \times 10^{-4}$ &
$0.716 \times 10^{-1}$ & $0.205\times 10^{-3}$ & 0.362 & $0.306\times 10^{-2}$ \\
$m_{\tilde{g}} \!=\!200$~GeV & $0.568\times 10^{-2}$  & $0.518 \times
10^{-4}$ & 0.335  & $ 0.242 \times 10^{-3}$ & 1.64 & $0.376 \times 10^{-2}$ \\
500~GeV & $0.157\times 10^{-2}$ & $0.247 \times 10^{-4}$ & $0.994\times
10^{-1}$& $0.234 \times 10^{-3}$ &
0.500 & $0.349 \times 10^{-2}$\\
1000~GeV & $0.447 \times 10^{-3}$ & $0.846 \times 10^{-5}$
& $0.297\times 10^{-1}$ & $0.124\times 10^{-3}$ & 0.153 & $0.187\times 10^{-2}$\\
\hline
\end{tabular}
\caption{The LO cross sections for $\tilde{b}_1\tilde{b}_1$ production at the Tevatron and 
         the LHC. The MSTW pdfs have been adopted and the scale has been set to
         the final-state sbottom mass.}
\label{tab:susydep_sbsb}
\end{small}
}

\end{appendix}

\clearpage

\pagebreak

\end{document}